%% file: main.tex
\documentclass[twocolumn,fleqn]{autart}
\usepackage[active]{srcltx}
\usepackage{cite}
\usepackage{array}

\usepackage[usenames, dvipsnames]{color}
\usepackage{verbatim}
\usepackage{booktabs}
\usepackage{units}
\usepackage{mathrsfs}
\usepackage{setspace}
\usepackage{upgreek}
\usepackage{graphicx, xcolor}
\usepackage{arydshln}
\usepackage{mathtools,cuted}

\usepackage{mathtools}
\usepackage{amssymb}
\usepackage{stmaryrd}
\usepackage{esint}
\usepackage{commath}
\usepackage{float}
\usepackage[toc,page]{appendix}
\usepackage{algpseudocode}
\usepackage{pgfplots}
\pgfplotsset{compat=newest}
\usepackage[inline]{enumitem}
\pgfplotsset{compat=1.18}
\pgfplotsset{/pgfplots/circular legend/.style={
        /pgfplots/legend image code/.code={%
            \fill[##1,/tikz/.cd, ] circle[radius=2pt];
            \draw[##1,/tikz/.cd,solid]
            (0cm,0.cm) -- (0cm,0cm);},},              
}

\DeclareMathOperator\conv{Conv}
\DeclareMathOperator\sign{sgn}

\DeclareMathOperator*{\argmin}{arg\,min}

\DeclareMathOperator{\Int}{Int}
\DeclareMathOperator{\diag}{diag}

\DeclareMathOperator\pr{\mathbf{Pr}}
\DeclareMathOperator\cbf{\scalemath{0.6}{\mathrm{CBF}}}

\allowdisplaybreaks
\newcommand\scalemath[2]{\scalebox{#1}{\mbox{\ensuremath{\displaystyle #2}}}}

\definecolor{color1}{RGB}{0,128,0}    
\definecolor{color2}{RGB}{255,0,0}    
\definecolor{color3}{RGB}{139,69,19}  
\definecolor{color4}{RGB}{0,0,255}    
\definecolor{color5}{RGB}{152,78,163} 
\definecolor{color6}{RGB}{255,165,0} 

\begin{document}
\begin{frontmatter}
	\title{Safe Output-Feedback Adaptive Optimal Control of Input-Constrained Control-Affine Nonlinear Systems}
	
	\thanks[footnoteinfo]{This research was supported in part by the Air Force Research Laboratory under contract numbers FA8651-24-1-0019 and FA8651-23-1-0006 and the Office of Naval Research under contract number N00014-21-1-2481. Any opinions, findings, or recommendations in this article are those of the author(s), and do not necessarily reflect the views of the sponsoring agencies.}
	
	\author[UF]{Tochukwu E. Ogri}\ead{tochukwu.ogri@ufl.edu},
 \author[UF]{Muzaffar Qureshi}\ead{muzaffar.qureshi@ufl.edu},
	\author[AFRL]{Zachary I. Bell}\ead{zachary.bell.10@us.af.mil},
	\author[UF]{Rushikesh Kamalapurkar}\ead{rkamalapurkar@ufl.edu}
	
	\address[UF]{Department of Mechanical and Aerospace Engineering, University of Florida, Gainesville, Florida, USA}
	\address[AFRL]{Air Force Research Laboratory, Florida, USA}
\begin{abstract} 
In this paper, a novel online, safe output-feedback, critic-only, adaptive optimal control framework is developed for safety-critical control of partially observable systems. The developed framework ensures system stability and safety, regardless of the lack of full-state measurements, while learning and implementing a near-optimal controller. The approach leverages linear matrix inequality-based observer design methods to efficiently search for observer gains for effective state estimation. Then, approximate dynamic programming is used to develop an approximate controller that uses simulated experiences to guarantee the safety and stability of the closed-loop system. Safety is enforced by adding a recentered robust Lyapunov-like barrier function to the cost function that effectively enforces safety constraints, even in the presence of state estimation errors. Lyapunov-based stability analysis is used to guarantee uniform ultimate boundedness of the trajectories of the closed-loop system and ensure safety. Simulation studies are performed to demonstrate the effectiveness of the developed method through two real-world safety-critical scenarios, specifically one ensuring that the state trajectories of a given system remain within a given set, and the other ensuring that the system avoids an obstacle.
\end{abstract}
\end{frontmatter}
\section{Introduction}

 In recent times, safe, learning-based optimal control (SLOC) has gained traction due to the need for robust control of safety-critical systems. While there are several SLOC strategies for nonlinear systems \cite{SCC.Chow.Nachum.ea2018, SCC.Yang.Vamvoudakis.ea2020,SCC.Deptula.Chen.ea2020,SCC.Greene.Deptula.ea2020,SCC.Mahmud.Nivison.ea2021,SCC.Cohen.Belta.ea2020, SCC.Marvi.Kiumarsi.ea2020, SCC.Almubarak.Theodorou.ea2021, SCC.Cohen.Belta.ea2023, SCC.Bandyopadhyay.Bhasin.ea2025}, most of these methods rely on full-state feedback, which is not always available in real-world systems. Without full state measurement, SLOC techniques like those in \cite{SCC.Cichosz1999, SCC.Wawrzynski2009, SCC.Zhang.Cui.ea2011, SCC.Adam.Busoniu.ea2012, SCC.Cohen.Belta.ea2023, SCC.Greene.Deptula.ea2020, SCC.Deptula.Chen.ea2020,SCC.Mahmud.Nivison.ea2021,SCC.Cohen.Belta.ea2020, SCC.Marvi.Kiumarsi.ea2020, SCC.Yang.Liu.Huang.ea2013, SCC.Modares.Lewis.ea2014} could be implemented using a na\"{i}ve replacement of the actual state with state estimates, but the accompanying stability guarantees are no longer valid. Consequently, there is a need for adaptive output-feedback optimal control techniques for nonlinear systems that guarantee safety and stability regardless of limited state information from sensors.

\subsection*{Related Work On SLOC}
To achieve the safety of a dynamical system, a widely employed technique in the field of safe adaptive optimal control is to perform a nonlinear coordinate transformation, initially introduced in \cite{SCC.Graichen.Petit.ea2009} and further extended by \cite{SCC.Yang.Vamvoudakis.ea2020, SCC.Greene.Deptula.ea2020,SCC.Mahmud.Nivison.ea2021} for nonlinear control-affine systems. Barrier functions are employed to transform the constrained optimal control problem into an unconstrained one; however, the safety constraints they can handle are typically limited to box constraints. 

Approaches such as \cite{SCC.Cohen.Belta.ea2020, SCC.Marvi.Kiumarsi.ea2020}  utilize control barrier functions (CBFs), introduced in \cite{SCC.Wieland.Allgower.ea2007, SCC.Ames.Xu.ea2017,  SCC.Jankovic.ea2018} to solve the SLOC problem. 
A drawback of these techniques is that there is a chance that the trajectories of the nonlinear system converge to an undesired equilibrium point, as stated in \cite{SCC.Reis.Aguiar.ea2021} due to the added safety constraint. 
Results such as \cite{SCC.Cohen.Belta.ea2023, SCC.Cohen.Serlin.ea2023} avoid such behavior by decoupling the safety and learning objectives. In particular, safe control commands that are closest to a desired stabilizing controller are implemented to
guarantee safety in a minimally invasive fashion. The technique developed in this paper differs from \cite{SCC.Cohen.Belta.ea2023} due to its use of a recentered robust Lyapunov-like control barrier function (RLCBF) to promote safety by incorporating the barrier function into the cost function. Re-centering of the barrier function, first developed in \cite{SCC.Wills.Heath.ea2002}, guarantees system convergence to a desired equilibrium point regardless of safety objectives. 
The SLOC technique employed in this paper also differs from the strategies in \cite{SCC.Deptula.Chen.ea2020} and \cite{SCC.Cohen.Belta.ea2020}, where a user-defined smooth scheduling function is employed to selectively penalize trajectories near the boundary of the safe set, whereas our formulation imposes safety as a hard constraint rather than a soft penalty, ensuring safety without relying on trajectory-dependent penalization.

More recently, \cite{SCC.Almubarak.Theodorou.ea2021, SCC.Bandyopadhyay.Bhasin.ea2025} proposed a novel approach to SLOC by minimizing a Lagrangian that includes both the cost function and a user-defined barrier Lyapunov function (BLF) to enforce state constraints. Specifically, \cite{SCC.Almubarak.Theodorou.ea2021} minimizes the constrained generalized Hamilton-Jacobi-Bellman (HJB) equation using a modified Galerkin successive approximation approach to approximate the value function and synthesize the safe controller. Conversely, \cite{SCC.Bandyopadhyay.Bhasin.ea2025} introduced an Actor-Critic-Identifier-Lagrangian algorithm designed to learn optimal control policies online while ensuring safety.

Recent advances in multi-objective safety-critical control seek to balance safety with other performance objectives. The authors in \cite{SCC.Almubarak.Sadegh.ea2025, SCC.Almubarak.AL-Sunni.ea2025} achieve this through the use of barrier states (BaS) to embed safety conditions directly into control design, creating safety-embedded systems where stabilizing the augmented system ensures safe operation of the original system. The safety-embedded system framework in \cite{SCC.Almubarak.Sadegh.ea2025} allows the use of existing control techniques to synthesize safe controllers that enforce safety, handle input constraints, and guarantee input-to-state safe stability. In \cite{SCC.Almubarak.AL-Sunni.ea2025}, the BaS framework was combined with the State-Dependent Riccati Equation (SDRE) approach, yielding the BaS-SDRE method, which synthesizes safe, nearly optimal feedback laws in real time.

While effective, the existing SLOC methods described above utilize full-state feedback and cannot be trivially extended to systems where full state information is unavailable. For example, as demonstrated in Section~\ref{section:robustCBF}, the methods in \cite{SCC.Cohen.Belta.ea2023, SCC.Cohen.Serlin.ea2023} cannot ensure safety if na\"{i}vely implemented using state estimates. 

\subsection*{SLOC using output-feedback}

Early output-feedback optimal control methods developed in results such as \cite{SCC.Yang.Liu.Huang.ea2013,SCC.Yang.Liu.ea2014} assume that the pseudoinverse of the output matrix exists, which implicitly restricts the technique to systems where the number of outputs is larger than the number of states. More recent output-feedback SLOC techniques such as \cite{ SCC.Mahmud.Nivison.ea2021} also rely on structural restrictions imposed on the output matrix so that each output is a direct measurement of a state variable.

To avoid assumptions on the form or rank of the output matrix, our preliminary work in \cite{SCC.Ogri.Mahmud.ea2023, SCC.Ogri.Bell.ea2023} developed an observer for control-affine nonlinear systems using the multiplier matrix approach, a technique introduced in works such as \cite{SCC.Rajamani1998, SCC.Zemouche.Bara.ea2005, SCC.Phanomchoeng.Rajamani2010, SCC.Zemouche.Boutayeb.ea2013, SCC.Wang.Bevly.ea2014, SCC.Wang.Yang.ea2017, SCC.Rajamani.Zemouche.ea2020, SCC.Zemouche.Rajamani.ea2022, SCC.Mohite.Alma.ea2024}. For certain nonlinear systems, this approach guarantees asymptotic convergence of estimation errors under Lyapunov-derived conditions. For example, \cite{SCC.Rajamani1998} established sufficient stability conditions using eigenvalues and eigenvectors of the linear stability matrix for Lipschitz systems. Building on this, \cite{SCC.Zemouche.Bara.ea2005} reformulated the error dynamics as an LPV system via the differential mean value theorem (DMVT) and applied Lyapunov analysis to derive linear matrix inequalities (LMIs) for observer gains. Later works \cite{SCC.Phanomchoeng.Rajamani2010, SCC.Wang.Yang.ea2017} developed LMI-based designs for bounded Jacobian systems. In particular, \cite{SCC.Wang.Yang.ea2017} extended the Luenberger-type observer \cite{SCC.Arcak.Kokotovic2001a} by adding a third gain to cancel nonconvex terms in the LMI stability condition. Although LMI-based methods \cite{SCC.Wang.Bevly.ea2014, SCC.Wang.Yang.ea2017, SCC.Mohite.Alma.ea2024} ensure uniform asymptotic convergence, they remain restricted to specific classes of nonlinear systems.
More recently, \cite{SCC.Jeon.Chakrabarty.ea2024} exploited the combined benefits of matrix multipliers-based LMIs and NNs to develop an exponentially stable neuro-adaptive observer for state and nonlinear function estimation in Lipschitz systems with partially modeled process dynamics.

In our previous work, we developed novel observers for control-affine systems whose dynamics have bounded Jacobians. The observer developed in \cite{SCC.Ogri.Mahmud.ea2023} was integrated with an online critic-only adaptive control architecture to learn a controller that keeps an input-constrained nonlinear system stable during the learning phase. The architecture in \cite{SCC.Ogri.Mahmud.ea2023} differs from the existing neural network (NN)-based observers in literature like \cite{SCC.Kim.Lewis.ea1997,SCC.Abdollahi.Patel2006,SCC.Yang.Liu.Huang.ea2013,SCC.Yang.Liu.ea2014, SCC.Farza.Sboui.ea2010} whose convergence analysis relies solely on negative terms that result from a $\sigma-$modification-like term added to the weight update laws. In these results, similar to adaptive control, the convergence of the observer weights to their true values cannot be expected, and the convergence of state estimates to the true state is not robust to disturbances and approximation errors.

While our previous works \cite{SCC.Ogri.Mahmud.ea2023} and \cite{SCC.Ogri.Bell.ea2023} extend the LMI-based observer design to a broader class of nonlinear systems in the control-affine form, they do not address system safety. Incorporating state estimators into SLOC architectures is challenging, as estimation errors can propagate through value function updates, potentially destabilizing the learning process or resulting in suboptimal policies. The fact that safety constraints must be satisfied even in the presence of estimation errors necessitates robust formulations. While robustification of CBFs to ensure safety in the presence of estimation errors has been studied in results such as \cite{SCC.Agrawal.Panagou.ea2022, SCC.Cosner.Singletary.ea2021}, to the best of our knowledge, there are no existing output-feedback SLOC methods that incorporate observer-based state estimation with formal safety guarantees under input constraints. The development of a unified framework that couples observer design with robust CBF-based safety filters for safe learning of optimal controllers under partial observations and input constraints is thus well-motivated.

\subsection*{Contribution}
This paper focuses on extending our preliminary work in \cite{SCC.Ogri.Mahmud.ea2023} and \cite{SCC.Ogri.Bell.ea2023} to address the problem of safety guarantees in online nonlinear output-feedback optimal control. To estimate the unknown states of the nonlinear control-affine system required to compute the controller, we design a nonlinear observer and use Lyapunov-based analysis to develop LMIs that characterize the observer gain matrices. Due to the use of state estimates in the control law, CBFs fail to guarantee safety of the closed-loop system, as demonstrated in Figure~\ref{fig:demonstration} of Section~\ref{section:robustCBF}. In this paper, a robust control barrier function (RCBF) is developed to guarantee safety in the presence of state estimation errors using a time-varying robustifying term that relies on a shrinking bound on the state estimation error. 

The time-varying term avoids an overly conservative controller behavior, which is a common pitfall of many robust CBF-based controllers \cite{SCC.Cosner.Singletary.ea2021,SCC.Dean.Taylor.ea2021}. The key technical challenge in the use of RCBFs to solve SLOC problems is the fact that the incorporation of the RCBF constraint results in a time-varying optimization problem. Due to non-stationarity, such problems are difficult to solve online. In this paper, we address this technical challenge by utilizing the fact that the optimal control problem can be made time-invariant by using an augmented system state. We then develop a critic-only SLOC architecture to learn the optimal value function as a function of the augmented state to yield a safe and approximately optimal controller. Unlike actor-critic methods popular in the literature \cite{SCC.Kamalapurkar.Walters.ea2016, SCC.Kamalapurkar.Rosenfeld.ea2016}, the framework in this paper utilizes a critic-only structure to provide an approximate solution to the HJB equation, which requires identifying fewer free parameters. 

\subsection*{Organization}
 The rest of the paper is organized as follows: Section~\ref{section:problemFormulation} contains the problem formulation which introduces the system and some key assumptions about the system, Section~\ref{section:stateEstimator} presents the design of the state observer for the nonlinear output-feedback system with a linear measurement model, Section~\ref{section:robustCBF} introduces the concept of robust barrier functions and develops conditions sufficient to guarantee safety despite state measurement errors, Section~\ref{section:controlDesign} presents the control design using an observer-based SLOC method, Section~\ref{section:stabilityAnalysis} contains stability analysis of the developed control architecture, Section~\ref{section:simulation} presents simulation studies on two systems with distinct safety constraints and we compare the effectiveness of our approach against similar methods in the literature, and Section~\ref{section:conclusion} concludes the paper.

\subsection{Notation}
In this paper, 
the symbol $\| \cdot \|$ represents the $2-$norm for vectors and the induced $2-$norm for matrices. $\mathrm{I}_{n}$ denotes an $n \times n$ identity matrix and $0_n$ denotes the $n \times n$ zero matrix. For a square matrix $A \in \mathbb{R}^{n \times n}$, $\lambda_{\text{max}}(A)$ and $\lambda_{\text{min}}(A)$ denote its maximum and minimum eigenvalues, respectively; $A^{\top}$ denotes its transpose; $A > 0$ ($A < 0$) indicates that $A$ is a positive definite (negative definite) matrix; and $A \geq 0$ ($A \leq 0$) indicates that $A$ is a positive semi-definite (negative semi-definite) matrix. The total derivative $\frac{\partial f(x)}{\partial x}$ is denoted by $\nabla_{x} f(x)$ and the partial derivative $\frac{\partial f(x, y)}{\partial x}$ is denoted by $\nabla_x f(x, y)$. The convex hull of a set $\mathcal{O}$ is denoted as $\conv(\mathcal{O})$. The symbol $(\cdot)_{(\cdot)}$ denotes the elements of the function or variable $(\cdot)$ at the indices indicated by the subscript.

\section{Problem Formulation and Preliminaries}	
\label{section:problemFormulation}

\subsection{System Description}
Consider nonlinear dynamical systems of the form 
\begin{equation}\label{eq:dynamics_x}
\dot{x} = f(x)+g(x)u ,\quad y = Cx, 
\end{equation}
where $x \in \mathbb{R}^{n}$ is the system state, $u \in \mathbb{R}^{m}$ is the control input,  $C \in \mathbb{R}^{q \times n}$ is the output matrix, and $y \in \mathbb{R}^{q}$ is the measured output. The functions $f: \mathbb{R}^n \to \mathbb{R}^{n}$ and $g: \mathbb{R}^n \to \mathbb{R}^{n \times m}$ denote the drift vector and the control effectiveness matrix, respectively. 
 The following assumptions on the functions $f$ and $g$ are needed to facilitate the development and analysis of the method in this paper.
\begin{assum}\label{ass:jacobianbounds}
    The functions $f$ and $g$ are known, differentiable functions. Given any compact ball $\mathcal{X} \subset \mathbb{R}^n$ and a compact set \ $\mathcal{U}\subset \mathbb{R}^m$ the partial derivatives satisfy the element-wise bounds
    \begin{gather}
	(K_{f_{1}})_{i,j} \leq \nabla_{(x)_{j}}(f(x))_{i} \leq (K_{f_{2}})_{i,j}, \\ 
	(K_{g_{1}})_{i,j} \leq  [\nabla_{(x)_{j}}(g(x))_{i,k}]u_{k} \leq (K_{g_{2}})_{i,j},
\end{gather}
for all $x \in \mathcal{X} \oplus \mathcal{X}$, $u \in \mathcal{U}$, $i, j = 1, \ldots, n$, and $k = 1, \ldots, m$, where the operator $\oplus$ denotes the Minkowski sum. Furthermore, the function $f$ satisfies $f(0) = 0$.
\end{assum}

\begin{rem}
Conditions similar to those in Assumption~\ref{ass:jacobianbounds} are commonly required in several observer design schemes (see, e.g., \cite{SCC.Zemouche.Bara.ea2005,SCC.Wang.Yang.ea2017,SCC.Rajamani.Zemouche.ea2020, SCC.Zemouche.Rajamani.ea2022}). However, while Jacobian bounds are typically assumed to be global in those papers, here the bounds are assumed to be local, which is a less conservative assumption.
\end{rem}
To develop the notion of safety, consider the set $\mathcal{S} \subset \mathbb{R}^{n}$ defined as the zero super-level set of a continuously differentiable function $h: \mathbb{R}^{n} \to \mathbb{R}$ such that
\begin{align}
    \mathcal{S} = \{x \in \mathbb{R}^{n}: h(x)\geq 0\}\label{eq:safeSet1}, \\
    \partial\mathcal{S} = \{x \in \mathbb{R}^{n}: h(x) = 0\}, \\
    \Int\left(\mathcal{S}\right) = \{x \in \mathbb{R}^{n}: h(x) > 0\}, \label{eq:safeSet3}
\end{align}
 where $\partial\mathcal{S}$ and $\Int\left(\mathcal{S}\right)$ represent the boundary and interior of the set $\mathcal{S}$, respectively.

The objective is to design an observer to estimate the system state online, using input-output measurements, and to simultaneously synthesize and utilize a controller that minimizes the cost functional defined in \eqref{eq:costFunctional}, under the saturation constraint $|(u)_{k}(t)| \leq {\overline{u} } > 0 \text{ for all } t \geq t_{0} \text{ and } k = 1, \hdots, m$ (i.e., $u\in\mathcal{U}$ where $\mathcal{U}$ is the $m-$dimensional hypercube with side length $2\overline{u}$), and the safety constraint $x^{u}(t;t_{0}, x_{0}) \in \mathcal{S}$ for all $t \geq t_{0}$, and sufficiently small $x_{0}$ (see Theorem~\ref{thm:boundedAndSafe}), while ensuring local uniform ultimate boundedness of the trajectories of the closed-loop system. 

In the following section, a state observer inspired by the extended Luenberger observer \cite{SCC.Arcak.Kokotovic2001a, SCC.Wang.Yang.ea2017, SCC.Acikmese.Martin.2008, SCC.Quintana.Bernal.ea2021} is developed to generate estimates of $x$.

\section{Observer Design using Jacobian Bounds}
\label{section:stateEstimator}

The state observer is designed under the assumption that the system state remains within a compact convex set $\mathcal{X}$ and that the control input also remains within a compact set $\mathcal{U}$. CBFs and control saturation are employed in the following sections to ensure that the control signal remains in $\mathcal{U}$ and the system state remains in $\mathcal{X}$, thus ensuring that the Jacobian bounds in Assumption~\ref{ass:jacobianbounds} remain valid along the system trajectories.

\subsection{State Observer}
    Using the fact that $f(x) = K_{f_{1}}x + (f(x) -K_{f_{1}}x)$ and  $g(x) = K_{g_{1}}x + (g(x)-K_{g_{1}}x)$, the nonlinear dynamics in \eqref{eq:dynamics_x} can be expressed in the form
$
    \dot{x} =Ax +\mathcal{F}(x) + \mathcal{G}_{u}(x,u),
$
where $A \coloneqq {K}_{f_{1}}+{K}_{g_{1}} \in \mathbb{R}^{n \times n}$,
$\mathcal{F}(x) \coloneqq -{K}_{f_{1}}x+f(x)$, and $\mathcal{G}_{u}(x,u) \coloneqq -{K}_{g_{1}}x + \sum_{i=1}^{m} g_{i}(x)(u)_{i}$. Under Assumption~\ref{ass:jacobianbounds}, the derivatives of $\mathcal{F}$ and $\mathcal{G}$ satisfy the element-wise inequalities
\begin{align}
	0 & \leq \nabla_{(x)_{j}}(\mathcal{F}(x))_{i} \leq (K_{f_{2}})_{i,j}-(K_{f_{1}})_{i,j} \text{, and } \label{eq:augJacF} \\ 
	0 & \leq \nabla_{(x)_{j}}(\mathcal{G}_{u}(x,u))_{i} \leq (K_{g_{2}})_{i,j}-(K_{g_{1}})_{i,j},\label{eq:augJacG}
\end{align}
 where $i, j = 1, \hdots, n$. Since the Jacobian bounds specified in Assumption~\ref{ass:jacobianbounds} are local, even if the system state $x$ remains within the compact ball $\mathcal{X}$ for all $t \geq t_{0}$ through the use of CBFs, there is no guarantee that the state estimate $\hat{x} \in \mathbb{R}^{n}$ will remain in $\mathcal{X}$ for all $t \geq t_{0}$. Inspired by the results of \cite{SCC.Zemouche.Rajamani.ea2022}, a state observer which utilizes the well-known Hilbert projection theorem \cite{SCC.Mazumdar.ea1974} is designed as
\begingroup\medmuskip=0mu\begin{multline}\label{eq:observerdynamics_x}
    	\dot{\hat{x}} = f[\pr(\hat{x})+L_{1}\left(y-C\pr(\hat{x})\right)] \\ + g[\pr(\hat{x})+L_{2}\left(y-C\pr(\hat{x})\right)]u + L_{3}\left(y-C\pr(\hat{x})\right),
\end{multline}\endgroup
where $L_{1}, L_{2}$, and $L_{3} \in \mathbb{R}^{n \times q}$ are observer gains,  $L_{1}\left(y-C\pr(\hat{x})\right)$ and $L_{2}\left(y-C\pr(\hat{x})\right)$ are injection terms,  $L_{3}\left(y-C\pr(\hat{x})\right)$ is a linear correction term, and $\pr: \mathbb{R}^{n} \to \mathcal{X}$ is the Hilbert projection, which is defined as
$
    \pr(\hat{x}) = \argmin_{s \in \mathcal{X}} \left\|\hat{x} - s\right\|.
$
The following Hilbert projection theorem ensures the existence and uniqueness of $\pr(\hat{x})$.
\begin{thm}\cite[Hilbert projection theorem]{SCC.Mazumdar.ea1974}
    Let $\mathcal{X}$ be a closed convex subset of $\mathbb{R}^{n}$. For any $\hat{x} \in \mathbb{R}^{n}$, there exists a unique element $\pr(\hat{x}) \in \mathcal{X}$ such that
$
    \left\|\hat{x} - \pr(\hat{x})\right\| = \inf_{s \in \mathcal{X}} \left\|\hat{x} - s\right\|.
$
\end{thm}
Properties of $\pr(\hat{x})$, highlighted in \cite[Lemma~1]{SCC.Zemouche.Rajamani.ea2022}, are used in this development to guarantee that the local Jacobian bounds remain valid for all $t\geq t_{0}$, notwithstanding that the bounds on the Jacobian of the functions $f$ and $g$ are not global. 
The state estimation error is defined as $\tilde{x} \coloneqq x -\hat{x}$, and the estimation error dynamics can be expressed as
$
	\dot{\tilde{x}} = \left(A-L_{3}C\right)(\pr(x)-\pr(\hat{x})) + \mathcal{F}(\pr(x)) 
    +\mathcal{G}_{u}(\pr(x),u)-\mathcal{F}[\pr(\hat{x})+L_{1}C\left(\pr(x)-\pr(\hat{x})\right)] 
    -\mathcal{G}_{u}[\pr(\hat{x})+L_{2}C\left(\pr(x)-\pr(\hat{x})\right),u]. 
$ 
If $x \in \mathcal{X}$, then $Ax = A\pr(x)$, $y-C\pr(\hat{x}) = C(\pr(x)-\pr(\hat{x}))$, $\mathcal{F}(x) = \mathcal{F}(\pr(x))$ and $\mathcal{G}_{u}(x, u) = \mathcal{G}_{u}(\pr(x), u)$. Proposition~1 from \cite{SCC.Zemouche.Rajamani.ea2022} implies that there exists 
 a function $(\Phi)_{i, j}: \mathbb{R}^{n} \times \mathbb{R}^{n} \to \mathbb{R}$ such that for all $x \in \mathcal{X}$ and $\hat{x} \in \mathbb{R}^{n}$ 
 $
      \pr(x) - \pr(\hat{x})  =  \left[\sum_{i=1}^{n}\sum_{j=1}^{n} (\Phi (x, \hat{x}))_{i, j}(H)_{i, j}\right] (x - \hat{x}), 
 $ 
 and $ \vert\left(\Phi(x, \hat{x})\right)_{i, j}\vert \leq 1$, where $(H)_{i,j} \coloneqq \mathbf{e}(i)\mathbf{e}^{\top}(j)$ and $\mathbf{e}(i) \in \mathbb{R}^{n}$ denotes the $i$-th standard basis vector in $\mathbb{R}^{n}$.
Therefore, the observer error dynamics can be equivalently expressed as
\begin{equation}\label{eq:augError2}
     \dot{\tilde{x}} = \mathcal{A}_{c}(\theta)\tilde{x} +\Delta_{\mathcal{F}}(x,\hat{x}) + \Delta_{\mathcal{G}_{u}}(x,\hat{x},u),
\end{equation}
where $\mathcal{A}_{c}(\theta) \coloneqq  (A-L_{3}C)\theta$ with the matrix parameter $\theta(t) \coloneqq \sum_{i=1}^{n}\sum_{j=1}^{n} (\Phi (x, \hat{x}))_{i, j}(H)_{i, j}$ belonging to the set $\mathcal{O}$ which is defined as $\mathcal{O} \coloneqq \{M \in \mathbb{R}^{n \times n} : M = \sum_{i=1}^{n}\sum_{j=1}^{n} (\Phi (x, \hat{x}))_{i, j}(H)_{i, j}, \quad 0 \leq \left(\Phi(x, \hat{x})\right)_{i, j} \leq 1\}$, 
$\Delta_{\mathcal{F}}(x,\hat{x})  \coloneqq \mathcal{F}(\pr(x))
-\mathcal{F}[\pr(\hat{x})+L_{1}C\left(\pr(x)-\pr(\hat{x})\right)]$, and
$\Delta_{\mathcal{G}_{u}}(x, \hat{x}, u) \coloneqq \mathcal{G}_{u}(\pr(x),u)-\mathcal{G}_{u}[\hat{x}+L_{2}C\left(\pr(x)-\pr(\hat{x})\right),u]$.
    
\subsection{Formulation Of LMI Conditions}
{\small \begin{table*}[ht!]\footnotesize\label{tab:matrixM}
 \centering
\begingroup\medmuskip=0mu\thinmuskip=0mu\thickmuskip=0mu\begin{equation}\label{eq:matrixM}
   M \coloneqq \begin{bmatrix}
        A_{\theta}^{\top}P+PA_{\theta}-C_{\theta}^{\top} R^{\top}-RC_{\theta}+2\alpha P && \sqrt{2}P + \left(\mathrm{I}_{n}-L_{1}C\right)^{\top}\left(K_{{f}_{2}} - K_{{f}_{1}}\right)^{\top} + \left(\mathrm{I}_{n}-L_{2}C\right)^{\top}\left(K_{{g}_{2}} - K_{{g}_{1}}\right)^{\top}    \\ \sqrt{2}P + \left(K_{{f}_{2}} - K_{{f}_{1}}\right)\left(\mathrm{I}_{n}-L_{1}C\right) + \left(K_{{g}_{2}} - K_{{g}_{1}}\right)\left(\mathrm{I}_{n}-L_{2}C\right)  && -3\mathrm{I}_{n}
        \end{bmatrix}. 
    \end{equation}\endgroup
    \hrulefill 
\end{table*}}
The functions $\mathcal{F}$ and $\mathcal{G}_{u}$ can be equivalently be expressed as
$
    \mathcal{F}(x) = \sum_{i=1}^{n}\mathbf{e}(i)(\mathcal{F}(x))_{i},
$
 and
 $
 \mathcal{G}_{u}(x, u) = \sum_{i=1}^{n}\mathbf{e}(i)(\mathcal{G}_{u}(x, u))_{i}
 $, respectively. By the DMVT \cite[Theorem 2.3]{SCC.Zemouche.Bara.ea2005}, for all $x \in \mathcal{X}$ and $\hat{x} \in  \mathbb{R}^{n}$, there exist $(s_{f})_{i, j} \in \conv(\pr(x), \pr(\hat{x}) + L_{1}C(\pr(x)-\pr(\hat{x})))$ such that the difference function $\Delta_{\mathcal{F}}$ can be expressed as 
$
\Delta_{\mathcal{F}}(x,\hat{x}) = \left[\sum_{i=1}^{n}\sum_{j=1}^{n}(H)_{i,j} (\varphi_{f}(x, \hat{x}))_{i, j}\right] \left(\mathrm{I}_{n}-L_{1}C\right)(x-\hat{x})
$,  
 where the function  $(\varphi_{f})_{i, j}: \mathbb{R}^{n} \times \mathbb{R}^{n}  \to \mathbb{R}$ is defined as
\begingroup\medmuskip=0mu\thinmuskip=0mu\thickmuskip=0mu\begin{equation}\label{eq:fDMVT}
   \scalemath{0.76}{(\varphi_{f})_{i, j}(x, \hat{x}) \coloneqq  \begin{cases}
         0,  & \text{if } (x)_{j} = (\hat{x})_{j}, \\
         \frac{\pr((x)_{j}) - \pr((\hat{x})_{j})}{(x)_{j} - (\hat{x})_{j}}\nabla_{(x)_{j}}(\mathcal{F}((s_{f})_{i,j}))_{i}, & \text{if } (x)_{j} \neq (\hat{x})_{j}.
    \end{cases}}
\end{equation}\endgroup
 The subsequent development requires the property that for all $x \in \mathcal{X}$ and $\hat{x} \in \mathbb{R}^{n}$,
 \begin{gather}
    L_{1}C(\Pr(x) - \Pr(\hat{x}))  \in \mathcal{X} \oplus \mathcal{X} \text{ and }\nonumber\\
    L_{2}C(\Pr(x) - \Pr(\hat{x})) \in \mathcal{X} \oplus \mathcal{X}.\label{eq:ChiProperty}
  \end{gather}
  Since $\mathcal{X} \oplus \mathcal{X}$ is a ball, if the induced 2-norm of $L_{1} C$ and $L_{2} C$ is less than 1, then $L_{1} C$ and $L_{2} C$ map $\mathcal{X} \oplus \mathcal{X}$ to itself, and \eqref{eq:ChiProperty} follows. These norm constraints have been incorporated into the LMI condition developed in Theorem~\ref{thm:stateObserver} to ensure the validity of \eqref{eq:ChiProperty}, and as a result, the inequalities in \eqref{eq:fIneq} and \eqref{eq:gIneq}.

Under Assumption~\ref{ass:jacobianbounds}, the partial derivative $\nabla_{(x)_{j}}(\mathcal{F}((s_{f})_{i,j}))_{i}$ is bounded, element-wise, by the bounds in \eqref{eq:augJacF} (i.e $0 \leq \nabla_{(x)_{j}}(\mathcal{F}((s_{f})_{i,j}))_{i} \leq (K_{f_{2}})_{i,j}-(K_{f_{1}})_{i,j}, \forall i, j = 1, \hdots, n)$.\footnote{The notation $\pr((x)_{j})$ is a slight abuse of notation used to simplify the expression. To be precise, $\pr((x)_{j})$ should be considered as $\pr((x)_{j}\mathbf{e}(j))$ and $\pr((\hat{x})_{j})$ should be considered as $\pr((\hat{x})_{j}\mathbf{e}(j))$.} 
Similarly, by the DMVT, for all $x \in \mathcal{X}$ and $\hat{x} \in  \mathbb{R}^{n}$, there exist $(s_{g})_{i, j} \in \conv(\pr(x), \pr(\hat{x}) + L_{2}C(\pr(x)-\pr(\hat{x})))$ such that the difference function $\Delta_{\mathcal{G}_{u}}$ can be expressed as
$
\Delta_{\mathcal{G}_{u}}(x,\hat{x},u)  = \left[\sum_{i=1}^{n}\sum_{j=1}^{n}(H)_{i,j}(\varphi_{g}(x, \hat{x}, u))_{i, j}\right] 
\times \left(\mathrm{I}_{n}-L_{2}C\right)(x-\hat{x}),
$ 
where $(\varphi_{g})_{i, j}: \mathbb{R}^{n} \times \mathbb{R}^{n} \times \mathbb{R}^{m} \to \mathbb{R}$ is defined as
\begingroup\medmuskip=0mu\thinmuskip=0mu\thickmuskip=0mu\begin{equation}\label{eq:gDMVT}
   \scalemath{0.79}{(\varphi_{g})_{i, j} \coloneqq  \begin{cases}
         0,  & \text{if } (x)_{j} = (\hat{x})_{j}, \\
         \frac{\pr((x)_{j}) - \pr((\hat{x})_{j})}{(x)_{j} - (\hat{x})_{j}}\nabla_{(x)_{j}}(\mathcal{G}_{u}((s_{g})_{i,j}, u))_{i}, & \text{if } (x)_{j} \neq (\hat{x})_{j},
    \end{cases}}
\end{equation}\endgroup
and under Assumption~\ref{ass:jacobianbounds}, the partial derivative $\nabla_{(x)_{j}}(\mathcal{G}_{u}((s_{g})_{i,j}, u))_{i}$ is bounded element-wise by the bounds in \eqref{eq:augJacG}.
Using the monotonicity and $1$-Lipschitz property of the Hilbert projection \cite{SCC.Zemouche.Rajamani.ea2022}, $0 \leq \frac{\pr((x)_{j}) - \pr((\hat{x})_{j})}{(x)_{j} - (\hat{x})_{j}} \leq 1$. The representations in \eqref{eq:fDMVT} and \eqref{eq:gDMVT} both leverage the DMVT to ensure the existence of points $(s_{f})_{ij}$ and $(s_{g})_{ij}$, which lie along convex paths between $\pr(x)$ and affine perturbations of $\pr(\hat{x})$, determined by the matrices $C$, $L_{1}$, and $L_{2}$, at which the Jacobian of $\mathcal{F}$ and $\mathcal{G}_{u}$ are evaluated, respectively.

Using Assumption~\ref{ass:jacobianbounds}, the DMVT, and the bounds in \eqref{eq:augJacF}, the difference function $\Delta_{\mathcal{F}}$ is bounded as (cf. \cite{SCC.Ogri.Mahmud.ea2023})
\begingroup\medmuskip=0mu\begin{equation}\label{eq:fIneq}
 \mathbf{0}_{n}\leq \Delta_{\mathcal{F}}(x,\hat{x}) \leq \left(K_{{f}_{2}} - K_{{f}_{1}}\right)\left(\mathrm{I}_{n}-L_{1}C\right)\left(x-\hat{x}\right), 
\end{equation}\endgroup
 and similarly,
 using the bounds in \eqref{eq:augJacG}, the difference function $\Delta_{\mathcal{G}_{u}}$ is bounded as
\begingroup\medmuskip=0mu\thinmuskip=2.5mu\begin{equation}\label{eq:gIneq}
\mathbf{0}_{n}\leq \Delta_{\mathcal{G}_{u}}(x,\hat{x},u) \leq \left(K_{{g}_{2}} - K_{{g}_{1}}\right)\left(\mathrm{I}_{n}-L_{2}C\right)(x-\hat{x}).
\end{equation}\endgroup
In the following theorem, a bound on the Lie derivative of a candidate Lyapunov function along the flow of the observer error system in \eqref{eq:augError2} is derived. The bound is subsequently used in Theorem \ref{thm:boundedAndSafe} to establish local ultimate boundedness of the closed-loop system.
\begin{thm}\label{thm:stateObserver}
Let $V_{x}\left(\tilde{x}\right) \coloneqq \tilde{x}^{\top} P\tilde{x}$ be a candidate Lyapunov function where $P \in \mathbb{R}^{n \times n}$ is a symmetric positive definite  matrix. If
\begin{enumerate}[label={H\arabic*}:]
  \item\label{hyp:bounds} Assumption~\ref{ass:jacobianbounds} holds and 
  \item\label{hyp:observerGains} there exist a symmetric positive definite matrix $P \in \mathbb{R}^{n \times n}$, and observer gains $L_{1}$, $L_{2}$, and $R \in \mathbb{R}^{n \times q}$ that satisfy the constraints $\|L_{1}C\|\leq 1$, $\|L_{2}C\|\leq 1$, and the matrix inequality,
     \begin{equation}\label{eq:lmi}
       M < 0, \quad \forall \theta \in \mathcal{O},
     \end{equation}
     where $M$ is defined in \eqref{eq:matrixM} at the top of Page~\pageref{tab:matrixM}, $\alpha \in \mathbb{R}_{>0}$ is a constant learning rate, $A_{\theta} \coloneqq A\theta$ and $C_{\theta} \coloneqq C\theta$,
\end{enumerate}
then for all $\tilde{x} \in \mathbb{R}^{n}$, $x \in \mathcal{X}$, and $u\in\mathcal{U}$, the Lie derivative of $V_{x}$ along the flow of \eqref{eq:augError2} satisfies
\begin{equation}\label{eq:VeIneq}
\dot{V}_{x}\left(\tilde{x},x,u\right) \leq -2\alpha V_{x}\left(\tilde{x}\right).
\end{equation}
\end{thm}

\begin{pf}  
Note that $V_{x}$ satisfies the inequality
$
   \lambda_{\min}(P)\left\|\tilde{x}\right\|^{2} \leq V_{x}\left(\tilde{x}\right) \leq \lambda_{\max}(P)\left\|\tilde{x}\right\|^{2}. 
$ 
Taking the Lie derivative of $V_{x}$ along the flow of 
\eqref{eq:augError2} yields
$
\dot{V}_{x}\left(\tilde{x},x,u\right)  =  \tilde{x}^{\top}\left(\mathcal{A}_{c}^{\top}(\theta)P 
 + \mathcal{A}_{c}(\theta)P\right)\tilde{x} +\Delta_{\mathcal{F}}^{\top}(x,\hat{x})P\tilde{x} 
 + \tilde{x}^{\top}P\Delta_{\mathcal{F}}(x,\hat{x}) + \Delta_{\mathcal{G}_{u}}^{\top}(x,\hat{x},u)P\tilde{x} + \tilde{x}^{\top}P\Delta_{\mathcal{G}_{u}}(x,\hat{x},u).
$ 
 Using the Cauchy-Schwarz inequality and Young’s inequality, it can be concluded that
$
    \Delta_{\mathcal{F}}^{\top}(x,\hat{x})P\tilde{x} + \tilde{x}^{\top}P\Delta_{\mathcal{F}}(x,\hat{x}) \leq \left\|P\tilde{x}\right\|^{2} + \|\Delta_{\mathcal{F}}(x,\hat{x})\|^{2}.
$
\ By applying the bound from \eqref{eq:fIneq}, it can be concluded that for all $x \in \mathcal{X}$ and $\hat{x} \in \mathbb{R}^{n}$
\begin{multline}\label{eq:BoundOnF}
    \Delta_{\mathcal{F}}^{\top}(x,\hat{x})P\tilde{x} + \tilde{x}^{\top}P\Delta_{\mathcal{F}}(x,\hat{x}) \leq \tilde{x}^{\top}P^{2}\tilde{x} +  \tilde{x}^{\top}\left(\mathrm{I}_{n}-L_{1}C\right)^{\top} \\ \times \left(K_{{f}_{2}} - K_{{f}_{1}}\right)^{\top}\left(K_{{f}_{2}} - K_{{f}_{1}}\right)\left(\mathrm{I}_{n}-L_{1}C\right)\tilde{x}.
\end{multline}
Following the same steps, by using Cauchy-Schwarz,
Young’s inequality and applying the bound in \eqref{eq:gIneq}, it can be concluded that for all $x \in \mathcal{X}$, $\hat{x} \in \mathbb{R}^{n}$ and $u \in \mathcal{U}$,
\begin{multline}
    \Delta_{\mathcal{G}_{u}}^{\top}(x,\hat{x},u)P\tilde{x} + \tilde{x}^{\top}P\Delta_{\mathcal{G}_{u}}(x,\hat{x},u) 
    \leq \|P\tilde{x}\|^{2} \\\qquad + \|\Delta_{\mathcal{G}_{u}}(x,\hat{x}, u)\|^{2}\leq \tilde{x}^{\top}P^{2}\tilde{x} + \tilde{x}^{\top}\left(\mathrm{I}_{n}-L_{2}C\right)^{\top}\\ 
    \times \left(K_{{g}_{2}} - K_{{g}_{1}}\right)^{\top}\left(K_{{g}_{2}} - K_{{g}_{1}}\right)\left(\mathrm{I}_{n}-L_{2}C\right)\tilde{x}.
\end{multline}
Therefore, if the algebraic Ricatti inequality (ARI)  
 \begin{multline}\label{eq:ARI}
     \mathcal{A}_{c}^{\top}(\theta)P 
     + \mathcal{A}_{c}(\theta)P + 2P^{2} + \left(\mathrm{I}_{n}-L_{1}C\right)^{\top}\left(K_{{f}_{2}} - K_{{f}_{1}}\right)^{\top}\\\times \left(K_{{f}_{2}} - K_{{f}_{1}}\right)\left(\mathrm{I}_{n}-L_{1}C\right) + \left(\mathrm{I}_{n}-L_{2}C\right)^{\top}\left(K_{{g}_{2}} - K_{{g}_{1}}\right)^{\top}\\ \times\left(K_{{g}_{2}} - K_{{g}_{1}}\right)\left(\mathrm{I}_{n}-L_{2}C\right) < 0
 \end{multline}
 is satisfied, then the Lie derivative is negative definite.
 Splitting the ARI in \eqref{eq:ARI} into three inequalities and applying the Schur complement Lemma \cite{SCC.Boyd.Ghaoui1994} on each inequality, the first quadratic inequality, $\mathcal{A}_{c}^{\top}(\theta)P  + \mathcal{A}_{c}(\theta)P + 2P^{2} < 0$, can be equivalently expressed in matrix form as
 \begin{equation}\label{eq:lmi1}
 \begin{bmatrix} 
 \mathcal{A}_{c}^{\top}(\theta)P 
 + \mathcal{A}_{c}(\theta)P   && \sqrt{2}P \\ \sqrt{2}P && -\mathrm{I}_{n}
\end{bmatrix} < 0, 
\end{equation}
 the second inequality, $(\mathrm{I}_{n}-L_{1}C)^{\top}(K_{{f}_{2}} - K_{{f}_{1}})^{\top}(K_{{f}_{2}} - K_{{f}_{1}})(\mathrm{I}_{n}-L_{1}C) \leq 0$, can be expressed as    
\begingroup\medmuskip=0mu\thinmuskip=0mu\thickmuskip=0mu\begin{equation}\label{eq:lmi2}
     \scalebox{0.9}{$\begin{bmatrix}
      \mathbf{0}_{n} && \left(\mathrm{I}_{n}-L_{1}C\right)^{\top}\left(K_{{f}_{2}} - K_{{f}_{1}}\right)^{\top}
      \\ \left(K_{{f}_{2}} - K_{{f}_{1}}\right)\left(\mathrm{I}_{n}-L_{1}C\right) && -\mathrm{I}_{n}
  \end{bmatrix} \leq 0$}, 
 \end{equation}
 \endgroup
 and the third inequality, $(\mathrm{I}_{n}-L_{2}C)^{\top}(K_{{g}_{2}} - K_{{g}_{1}})^{\top}(K_{{g}_{2}} - K_{{g}_{1}})(\mathrm{I}_{n}-L_{2}C) \leq 0$, can be expressed as
\begingroup\medmuskip=0mu\thinmuskip=0mu\thickmuskip=0mu\begin{equation}\label{eq:lmi3}
\scalebox{0.9}{$\begin{bmatrix}
      \mathbf{0}_{n} && \left(\mathrm{I}_{n}-L_{2}C\right)^{\top}\left(K_{{g}_{2}} - K_{{g}_{1}}\right)^{\top}
      \\ \left(K_{{g}_{2}} - K_{{g}_{1}}\right)\left(\mathrm{I}_{n}-L_{2}C\right) && -\mathrm{I}_{n}
  \end{bmatrix} \leq 0$}.
  \end{equation}
\endgroup  
Combining the matrix inequalities in \eqref{eq:lmi1}, \eqref{eq:lmi2}, and \eqref{eq:lmi3}, it can be concluded that if the LMI in \eqref{eq:lmi} is satisfied for some constant $\alpha \in \mathbb{R}_{>0}$, then the Lie derivative is bounded as described in \eqref{eq:VeIneq}.
\end{pf}
\begin{rem}
    The observer gain $L_{3}$ can be obtained via the typical variable substitution $L_{3} = P^{-1}R$.
\end{rem}

\section{Safe control using RCBFs}\label{section:robustCBF}
\begin{figure}
        \centering
        \input{figures/Cohen2020}
	\caption{Trajectories of the state and estimated state for the system in \eqref{eq:simDyn} when the SLOC framework from \cite{SCC.Cohen.Belta.ea2020} is used to solve the obstacle avoidance problem in Section~\ref{subsection:simStudy2}. It can be observed that while the state estimate trajectory $\hat{x}(t)$ remains outside the obstacle boundary, the actual state trajectory $x(t)$ breaches it. The failure of the actual system at avoiding the obstacle when using a controller with state estimates highlights the limitations of relying solely on CBFs to guarantee the safety of an output feedback nonlinear system without the
    augmentation of the CBF with a robustifying term.}
		\label{fig:demonstration}
\end{figure}
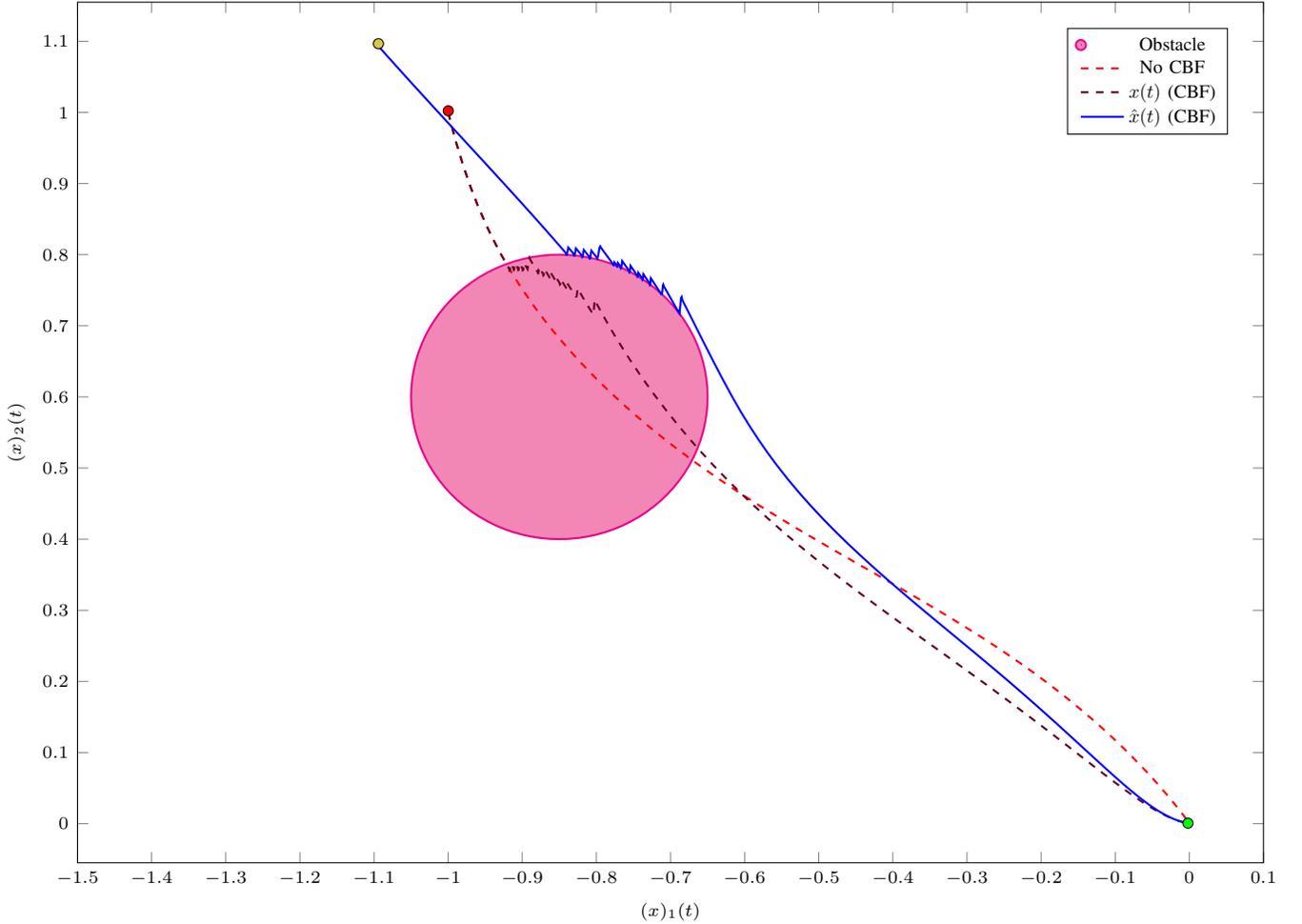

The state-feedback SLOC techniques developed by \cite{SCC.Cohen.Belta.ea2020} and \cite{SCC.Marvi.Kiumarsi.ea2020}, which embed the CBF in the cost function of the optimal control problem to enforce safety constraints, fail in the output-feedback case when implemented via na\"{i}ve certainty equivalence without introducing a robustifying term. Figure~\ref{fig:demonstration} illustrates this failure using the system from the simulation study in Section~\ref{section:simulation}. 

Inspired by RCBF techniques utilized in \cite{SCC.Jankovic.ea2018, SCC.Takano.Yamakita.ea2018, SCC.Kolathaya.Ames.ea2019, SCC.Dean.Taylor.ea2021, SCC.Agrawal.Panagou.ea2022, SCC.Lopez.Slotine.ea2020} to ensure safety in the presence of state estimation errors, this section introduces RLCBFs which will be utilized in this paper to achieve safety objectives. The purpose of this section is to motivate the design of the RLCBF introduced in \eqref{eq:controlBarrier} and to provide the rationale behind its selection. It should be noted that the stability results presented herein are not directly utilized in the control design; they only justify the choice of the final RLCBF used in the control design, as detailed in Section~\ref{section:controlDesign}.

 \subsection{Control Barrier Functions}

  The following definitions from \cite{SCC.Ames.Xu.ea2017} formalize the concept of  CBFs, which are employed to enforce the safety constraints.

\begin{defn}\label{defn:cbf}
\cite{SCC.Ames.Xu.ea2017} Given a set $\mathcal{S} \subset \mathcal{X} \subset \mathbb{R}^{n}$ as defined in \eqref{eq:safeSet1}--\eqref{eq:safeSet3} and the set \ $\mathcal{U} \subset {R}^{m}$, a continuously differentiable function $h: \mathcal{X} \to \mathbb{R}$ is a CBF for \eqref{eq:dynamics_x}, if there exist a class $\mathcal{K}$ function $\nu$ such that,
$
    \sup_{u \in \mathcal{U}}\nabla_{x}h(x)\left(f(x) + g(x)u\right) \geq -\nu\left(h(x)\right), \forall x \in \mathcal{S}.
$
\end{defn} 
The following theorem from \cite{SCC.Ames.Xu.ea2017}, stated here for completeness, establishes conditions that must be satisfied for the existence of control policies that guarantee safety.
\begin{thm}\label{thm:forwardInvariantC}
    \cite[Theorem~1]{SCC.Ames.Xu.ea2017} Given the sets $\mathcal{S} \subset \mathcal{X} \subset {R}^{n}$ and \ $\mathcal{U} \subset {R}^{m}$, along with a continuously differentiable CBF $h: \mathcal{X} \to \mathbb{R}$ for the system in \eqref{eq:dynamics_x} such that $\nabla h(x) \neq 0$ for all $x \in \partial\mathcal{S}$ and a set defined as
{\medmuskip=0mu\thinmuskip=0mu\thickmuskip=1mu\begin{equation}\label{eq:kCBF}
    \mathcal{K}_{\cbf} \coloneqq \left\{u \in \mathcal{U}: \nabla_{x} h(x)\left(f(x) + g(x)u\right) \geq -\nu\left(h(x)\right)\right\},
\end{equation}}any control policy $\pi_{\phi}: \mathcal{X} \times \mathbb{R}_{\geq 0}  \to \mathcal{U}$ that is locally Lipschitz continuous in $x$, piecewise continuous in $t$, and satisfies $\pi_{\phi}(x, t) \in \mathcal{K}_{\cbf}$ for all $x \in \mathcal{X}$ and $t \geq t_{0}$, guarantees forward invariance of the set $\mathcal{S}$.
\end{thm}

\subsection{Safety using RCBFs}

Since state estimates are used for feedback instead of the true state, the performance of CBFs in guaranteeing the safety of the system in \eqref{eq:dynamics_x} may be degraded. A RCBF \cite{SCC.Lopez.Slotine.ea2020} can provide robustness despite the error resulting from the difference between the exact and approximate controller.

Even if a given control policy satisfies $\pi_{\phi}(x,t) \in \mathcal{K}_{\cbf}$ for all $(x,t) \in \mathcal{X} \times \mathbb{R}_{\geq 0}$, the implemented controller relies on the estimated state, $u(t) = \pi_{\phi}(\hat{x},t)$. Since $\hat{x} \neq x$ in general, the resulting control input may not guarantee the safety of the actual system \eqref{eq:dynamics_x}. By using the observer designed in Section~\ref{section:stateEstimator} and the bound on the state estimation error in \eqref{eq:stateErrorBound}, an RCBF can be developed to guarantee the safety of the controller $u(t) = \pi_{\phi}(\hat{x},t)$ regardless of the state estimation error. To that end, let $\hat{\mathcal{S}}$ denote a set of state estimates defined as
\begin{equation}\label{eq:safeSetEstimate}
    \hat{\mathcal{S}}(t) \coloneqq \left\{\hat{x} \in \mathbb{R}^{n} : h_{r}(\hat{x}, t) \geq 0 \right\}, 
\end{equation}
where the function $h_{r} \in \mathbb{R}^{n} \times \mathbb{R}_{\geq 0} \to \mathbb{R}$ is defined as $h_{r}(\hat{x}, t) = h(\hat{x}) - \ell\xi(t)$, where $\ell \in \mathbb{R}_{>0}$ is a Lipschitz constant that satisfies 
$
    \vert h(x)-h(\hat{x})\vert \leq \ell\left\|x-\hat{x}\right\|, \quad \forall x, \ \hat{x} \in \mathbb{R}^{n},
$
where $\xi: \mathbb{R}_{\geq 0} \to \mathbb{R}$ is a non-increasing robustifying term defined as 
\begin{equation}\label{eq:compensationTerm}
    \xi(t) \coloneqq \left(\sqrt{\frac{\lambda_{\max}(P)}{\lambda_{\min}(P)}}\right) \epsilon_{0}\exp({-\alpha t}),
\end{equation}
where $\epsilon_{0} \in \mathbb{R}_{>0}$ is a constant such that $\left\|x(t_{0})-\hat{x}(t_{0})\right\| \leq \epsilon_{0}$. The robustifying term in \eqref{eq:compensationTerm} depends on the symmetric positive definite matrix $P$, which is obtained by solving the LMI in \eqref{eq:lmi}. The robustifying term in \eqref{eq:compensationTerm} is inspired by approaches such as \cite{SCC.Jankovic.ea2018, SCC.Lopez.Slotine.ea2020, SCC.Agrawal.Panagou.ea2022, SCC.Wang.Xu.ea2022}, but it is redesigned to integrate with our observer design. The robustifying term is obtained using the fact that if the control signal $u(\cdot)$ and the corresponding trajectory $x(\cdot)$ of the system in \eqref{eq:dynamics_x} satisfy $x(t) \in \mathcal{X}$ and $u(t) \in \mathcal{U}$ for all $t \geq t_{0}$ such that the bound in \eqref{eq:VeIneq} holds, then the Comparison Lemma \cite[Lemma~3.4]{SCC.Khalil2002} can be invoked to develop the bound
\begin{equation}\label{eq:stateErrorBound}
        \left\|\tilde{x}(t)\right\| \leq \xi(t), \quad \forall t \geq t_{0}.
\end{equation}
The following definition formalizes the concept of an RCBF for nonlinear systems with partial state measurement.
\begin{defn}\label{defn:rCBF}
    Given a set $\mathcal{S}\subset \mathbb{R}^{n}$ defined in \eqref{eq:safeSetEstimate}, a continuously differentiable function $h_{r}: \mathcal{X} \times \mathbb{R}_{\geq 0} \to \mathbb{R}$ is a RCBF for \eqref{eq:dynamics_x} with an observer\eqref{eq:observerdynamics_x} of known error bound \eqref{eq:stateErrorBound}, if there exist a class $\mathcal{K}$ function $\nu$ such that,
   $
    \sup_{u \in \mathcal{U}} 
     \nabla_{x} h_{r}(x, t)\left(f(x) + g(x)u\right) + \nabla_{t} h_{r}(x, t) 
     \geq  \nu\left(h_{r}(x, t_{0})\right), 
$ 
for all $x \in \mathcal{S}$ and $t \geq t_{0}$.
\end{defn}
The RCBF in Definition~\ref{defn:rCBF} is motivated by similar definitions in \cite{SCC.Jankovic.ea2018, SCC.Wang.Xu.ea2022, SCC.Agrawal.Panagou.ea2022}.

\begin{lem}\label{lem:safeBackupSet}
Given the set $\mathcal{S} \subset \mathcal{X} \subset \mathbb{R}^{n}$ defined by an RCBF $h_{r}: \mathbb{R}^{n} \times \mathbb{R}_{\geq 0} \to \mathbb{R}$ that satisfies Definition~\ref{defn:rCBF} and control set $\mathcal{U} \subset \mathbb{R}^{m}$, if Assumption~\ref{ass:jacobianbounds} holds, Hypothesis \ref{hyp:observerGains} of Theorem \ref{thm:stateObserver} holds, and $x(t) \in \mathcal{S},\forall t\geq t_0$, then any control policy $u = \pi_{\phi}(\hat{x},  t)$ that ensures $\hat{x}_{0} \in \hat{\mathcal{S}}(t) \implies h_{r}(\hat{x},t) \geq 0$ for all $\hat{x} \in \mathbb{R}^{n}$ and $t \geq t_{0}$ guarantees the safety of the system in \eqref{eq:dynamics_x} with respect to $\mathcal{S}$.
\end{lem}
\begin{pf}
     Let $u = \pi_{\phi}(\hat{x}, t)$ be a control policy such that $\hat{x}_{0} \in \hat{\mathcal{S}}$ implies $h_{r}(\hat{x},t) \geq 0$ for all $\hat{x} \in \mathbb{R}^{n}$ and $t \geq t_{0}$. By the Lipschitz continuity of $h$ and using the triangle inequality
     $
       \vert h(x) - h(\hat{x}) \vert \leq \ell\|x-\hat{x}\| \implies h(\hat{x}) - \ell\|x-\hat{x}\| 
       \leq h(x) \leq  h(\hat{x}) + \ell\|x-\hat{x}\|. 
    $
    \ Since $h_{r}(\hat{x},t) \geq 0$ and $\ell\|x-\hat{x}\| \leq \ell\xi(t)$ for all $x, \hat{x} \in \mathbb{R}^{n}$, and $t \geq t_{0}$, it can be concluded that the inequality
$
    0 \leq h(\hat{x}) - \ell\xi(t) \leq  \ell\|x-\hat{x}\| \leq h(x)
$
 holds for all $t \geq t_{0}$. As a result, $h(x) \geq 0$ for all $t \geq t_{0}$, which implies that $x \in \mathcal{S}$ from \eqref{eq:safeSet1} and thus, the system in \eqref{eq:dynamics_x} safe.
\end{pf}
In the following section, a safe control policy that utilizes RCBFs in the cost function of the output-feedback optimal control problem will be designed, so that trajectories of the system in \eqref{eq:dynamics_x} remain in a given set $\mathcal{S}$ despite the lack of full state measurement.

\section{Control Design}
\label{section:controlDesign}

\subsection{ADP-Based Safe and Stabilizing Optimal Control}
The developed technique uses a cost function that includes the RCBF to meet the safety constraints. Since the RCBF depends on the robustifying term $\xi$ in \eqref{eq:compensationTerm}, so does the value function. To address this dependency, a state augmentation technique similar to \cite{SCC.Kamalapurkar.Dinh.ea2015} is used. Let $\zeta \coloneqq [x^{\top}, \xi]^{\top} \in \mathbb{R}^{n + 1}$ be a concatenated vector such that the augmented dynamics are given by
\begin{equation}\label{eq:dynamics_zeta}
    \dot{\zeta} = F(\zeta) + G(\zeta)u,
\end{equation}
where the augmented functions $F: \mathbb{R}^{n+1} \to \mathbb{R}^{n+1}$ and $G: \mathbb{R}^{n+1} \to \mathbb{R}^{(n+1) \times m}$ are defined as $F(\zeta) \coloneqq \begin{bmatrix} f^{\top}(x), & -\alpha\xi\end{bmatrix}^{\top}$ and $G(\zeta) {\coloneqq}\begin{bmatrix} g^{\top}(x), & 0_{m \times 1} \end{bmatrix}^{\top}$, respectively. 
   
The optimal control problem is formulated using approximate dynamic programming (ADP)-based design techniques to develop a control signal, $u: \mathbb{R}_{\geq 0} \to \mathbb{R}^{m}$, online, that minimizes the cost functional defined as
\begin{equation} \label{eq:costFunctional}
J\left(\zeta,u(\cdot)\right)  \coloneqq  \int_{t}^\infty  r\left(\zeta^{u}(t_{0};\tau, \zeta_{0}), u(\tau)\right) d\tau,
\end{equation}
over the set of piecewise continuous functions $u: \mathbb{R}_{\geq 0} \to \mathcal{U}$, 
where $\zeta^{u}(t_{0};t,\zeta_{0})$ is a solution of \eqref{eq:dynamics_zeta} under control signal $u(\cdot)$ starting from initial condition $\zeta(t_{0}) = \zeta_{0} \in \mathbb{R}^{n+1}$, and $r: \mathbb{R}^{n + 1} \times \mathbb{R}^{m} \to \mathbb{R}$ is the instantaneous cost defined as
\begin{equation}\label{eq:costFunction}
r\left(\zeta,u\right) \coloneqq \overline{Q}(\zeta) + U\left(u\right) + B_{r}(\zeta)
\end{equation}
where the function $\overline{Q}: \mathbb{R}^{n + 1} \to \mathbb{R}$ is defined as 
$\overline{Q}([x^{\top},\xi]^{\top}) \coloneqq Q(x), \forall x \in \mathcal{X}$ and $\forall \xi \in \mathbb{R}$ with $Q: \mathcal{X} \to \mathbb{R}$ representing  a continuous positive definite
function. By \cite[Lemma~4.3]{SCC.Khalil2002}, $Q$ satisfies
$
    \underline{q}\left(\|x\|\right) \leq Q(x) \leq \overline{q}\left(\|x\|\right),
$ 
where $\underline{q}, \overline{q} : \mathbb{R}_{\geq 0} \to \mathbb{R}_{\geq 0}$ are class $\mathcal{K}$ functions. The function $U:\mathbb{R}^{m}\to\mathbb{R}$, introduced to enforce the saturation constraint on the control input, is defined as
\begin{equation} \label{eq:controlIntegrandfunction}
    U\left(u\right) \coloneqq 2\int_{0}^{u} \left({\overline{u} }\tanh^{-1}\left(\frac{\upsilon}{\overline{u} }\right)\right)^{\top} Rd\upsilon,
\end{equation}
where $R \coloneqq \diag(r_{1},\hdots, r_m)$. 
In this development, the function
 $B_{r}$ is selected as an RLCBF of the form
\begin{equation}\label{eq:controlBarrier}
    B_{r}(\zeta) \coloneqq \Big(b_{r}(\zeta)-b_{r}\left(0\right)\Big)^{2},
\end{equation}
where $b_{r}(\zeta) \coloneqq -\ln\left(\frac{\kappa h_{r}(\zeta)}{\kappa h_{r}(\zeta)+1}\right)$ and $\kappa  \in \mathbb{R}_{>0}$ is a user-defined constant that determines the magnitude of the barrier penalty as the trajectories approach the boundary of $\mathcal{S}$. Assuming the optimal controller exists, the optimal value function, $V^{*}: \mathbb{R}^{n + 1}  \to \mathbb{R}$ can be expressed as  
\begin{equation} \label{eq:valuefunction}
V^{*}(\zeta) \coloneqq \scalemath{0.8}{\min_{u(\tau) \in \mathcal{U}, \tau \in \mathbb{R}_{\geq t}}}\int_{t}^\infty  r\left(\zeta^{u}(t_{0};\tau, \zeta_{0}), u(\tau)\right)  d\tau.
\end{equation}
Assuming that the optimal value function is continuously differentiable, it can be shown to be the unique positive definite solution of the HJB equation, \cite[Theorem 1.5]{SCC.Kamalapurkar.Walters.ea2018},
\begin{equation}\label{eq:HJB} 
\min_{u\in \mathcal{U}} \Bigl(\nabla_{\zeta} V(\zeta)\left(F(\zeta)+G(\zeta)u\right) + r\left(\zeta,u\right) \Bigr) = 0.
\end{equation}
Therefore, the optimal stabilizing controller is given by the feedback policy $u\left(t\right) = u^{*}(\zeta^{u}(t_{0};\tau, \zeta_{0}))$,  where $u^{*}$ is given by 
\begin{equation}\label{eq:optimalcontrol}
    u^{*}(\zeta) \coloneqq -{\overline{u} }\tanh\left(D^{*}(\zeta)\right),
\end{equation}
where $D^{*}(\zeta) \coloneqq \frac{R^{-1}G^{\top}(\zeta)}{2\overline{u} } \nabla_{\zeta}(V^{*}(\zeta))^{\top} \in\mathbb{R}^{m}$. When the true states $x$ are available, the robustifying term in \eqref{eq:compensationTerm} vanishes because $\epsilon_{0} \equiv 0$ when $x = \hat{x}$. In this case, the RLCBF-based SLOC in \eqref{eq:optimalcontrol} reduces to the input-constrained version of existing CBF-based SLOCs (e.g., \cite{SCC.Cohen.Belta.ea2020,SCC.Marvi.Kiumarsi.ea2020}).

\subsection{Value Function Approximation }

Solving the HJB equation in \eqref{eq:HJB} is generally infeasible for nonlinear systems; hence, to find an approximate solution, estimates of the value function and the control policy are introduced.  Let $\mathcal{B}(0, \chi)$ be a closed ball of radius $\chi$ containing the origin, where $\chi$, defined as $\chi \coloneqq \left(\sqrt{\frac{\lambda_{\max}(P)}{\lambda_{\min}(P)}}\right) \epsilon_{0}$, is obtained from the bound on the state estimation error in \eqref{eq:stateErrorBound}. The unknown difference between the optimal value function $V^{*}$ and the barrier function $B_{r}$ can be expressed using a NN for local parametric approximation over the compact set $\Omega \subset \left(\mathcal{X} \oplus \mathcal{B}(0, \chi)\right)  \times \mathcal{B}(0, \chi)$ containing the origin as
\begin{equation} \label{eq:optimalV}
V^{*}(\zeta) -B_{r}(\zeta) = W^{\top} \phi(\zeta)+\epsilon(\zeta), \quad \forall \zeta \in \Omega,
\end{equation}
where $W\in\mathbb{R}^{L}$ is an unknown vector of bounded weights, $\phi: \mathbb{R}^{n+1} \to \mathbb{R}^{L}$ is a vector of continuously differentiable nonlinear activation functions that satisfy $\phi\left(0\right)=0$ and $\nabla_{\zeta} \phi \left(0\right)=0$, $L$ is the number of basis functions, and $\epsilon: \mathbb{R}^{n+1}\to\mathbb{R}$ is the reconstruction error. Invoking Stone-Weierstrass Theorem\cite[Theorem 1.5]{SCC.Sauvigny2012}, the activation functions $\phi$ can be selected so that the weights and the approximation errors satisfy $\sup_{\zeta \in \Omega}\|W\| \leq \overline{W}$, $ \sup_{\zeta \in \Omega}\|\phi(\cdot)\| \leq \overline\phi$, $ \sup_{\zeta \in \Omega}\|\nabla_{\zeta}\phi(\cdot)\| \leq \overline{\nabla\phi}, \sup_{\zeta \in \Omega}\|\epsilon(\cdot)\| \leq \overline{\epsilon}$ and $\sup_{\zeta \in \Omega}\|\nabla_{\zeta}\epsilon(\cdot)\| \leq \overline{\nabla\epsilon}$, where the notation $\overline{(\cdot)} \coloneqq \sup_{\zeta \in \Omega} \|(\cdot)\|$ is a positive constant.

 Since the actual state $\zeta$ is unknown and the ideal weights $W$ are unknown, let $\hat{\zeta} \coloneqq \begin{bmatrix}
    \hat{x}^{\top}, \xi
\end{bmatrix} ^{\top} \in \Omega$ be an estimate of the concatenated state $\zeta$ and let the estimates of the optimal value function and the optimal controller denoted as $\hat{V}: \mathbb{R}^{n+1} \times\mathbb{R}^{L}\to\mathbb{R}$ and $\hat{u}: \mathbb{R}^{n+1} \times\mathbb{R}^{L}\to\mathbb{R}^{m}$, respectively, be defined as
\begin{equation}\label{eq:VAppp}
\hat{V}(\hat{\zeta},\hat{W}_{c})\coloneqq\hat{W}_{c}^{\top} \phi(\hat{\zeta}) + B_{r}(\hat{\zeta}) \quad \text{and}
\end{equation} 
\begin{equation}\label{eq:uApp}
\hat{u}(\hat{\zeta},\hat{W}_{c}) \coloneqq-{\overline{u} }\tanh\left(\hat{D}(\hat{\zeta},\hat{W}_{c})\right),
\end{equation}
for all $\hat{\zeta} \in \Omega$, where $\hat{D}(\hat{\zeta},\hat{W}_{c}) \coloneqq \frac{R^{-1}G(\hat{\zeta})^{\top}}{2\overline{u}} (\nabla_{\hat{\zeta}}\phi(\hat{\zeta})^{\top} \hat{W}_{c} + \nabla_{\hat{\zeta}}B_{r}^{\top}(\hat{\zeta})) \in \mathbb{R}^{m}$ and  $\hat{W}_{c}\in\mathbb{R}^{L}$ are estimates of the ideal weights $W$. 

\subsection{Bellman Error and Simulation of Experience}
By substituting the approximations of the optimal value function and optimal controller from \eqref{eq:VAppp} and \eqref{eq:uApp}, respectively, into the HJB equation in \eqref{eq:HJB}, the residual term $\hat{\delta}: \mathbb{R}^{n+1} \times \mathbb{R}^{L} \to \mathbb{R}$, referred to as the Bellman error (BE), is obtained as
\begin{multline}\label{eq:BE1}
\hat{\delta}(\hat{\zeta},\hat{W}_{c}) =  \nabla_{\hat{\zeta}}\hat{V}(\hat{\zeta},\hat{W}_{c})\left(F(\hat{\zeta}) + G(\hat{\zeta})\hat{u}(\hat{\zeta},\hat{W}_{c})\right) \\ + \overline{Q}(\hat{\zeta}) + U\left(\hat{u}\right) + B_{r}(\hat{\zeta}).
\end{multline} 

To accurately approximate the value function, online SLOC methods require the persistence of excitation (PE) condition \cite{SCC.Modares.Lewis.ea2013, SCC.Kamalapurkar.Rosenfeld.ea2016}, which is difficult to guarantee in practice. However, through BE extrapolation for excitation via simulation, stability and convergence of online SLOC can be established \cite{SCC.Kamalapurkar.Rosenfeld.ea2016}.
To simulate experience using BE extrapolation,  select a set of trajectories $\left\{ \zeta_{k}: \mathbb{R}_{\geq 0} \to \Omega: k=1,\cdots, N\right\}$ and extrapolate the BE along these trajectories to yield the extrapolated BEs, $\hat{\delta}_{k}: \mathbb{R}^{n+1}  \times \mathbb{R}^{L} \to \mathbb{R}$, given by
 \begin{multline} \label{eq:BE1i}
    \hat{\delta}_{k}(\zeta_{k},\hat{W}_{c}) \coloneqq \nabla_{\zeta_{k}}\hat{V}(\zeta_{k},\hat{W}_{c})\left(F(\zeta_{k})+\hat{G}(\zeta_{k})\hat{u}(\zeta_{k},\hat{W}_{c})\right)\\+ \overline{Q}(\zeta_{k}) + U\left(\hat{u}\right) + B_{r}(\zeta_{k}).
\end{multline}
The control objective is achieved by updating the critic NN weights $\hat{W}_{c}$ online to minimize the BE using an adaptive update law developed from the subsequent stability analysis in Section~\ref{section:stabilityAnalysis}. 

\subsection{Update laws for Critic weights}
To guarantee that the estimated value function weights, $\hat{W}_{c}$, converge to their ideal values in (\ref{eq:optimalV}), the  estimated value function weights are updated based on the stability analysis in Section~\ref{section:stabilityAnalysis} as
\begin{gather}
    \dot{\hat{W}}_{c} = - \frac{k_{c}}{N}\Gamma\sum_{k=1}^{N}\frac{\omega_{k}}{\rho_{k}}\hat{\delta}_{k},\;
    \dot{\Gamma} = \beta\Gamma- \frac{k_{c}}{N}\Gamma\sum_{k=1}^{N}\frac{\omega_{k}\omega_{k}^{\top}}{\rho_{k}^{2}}\Gamma,\label{eq:weightsUpdate}
\end{gather}
with $\Gamma\left(t_{0}\right)=\Gamma_{0}$, where $\Gamma: \mathbb{R}_{\geq 0} \to \mathbb{R}^{L\times L}$
is a time-varying least-squares gain matrix, $\omega_{k} \coloneqq \nabla_{\zeta_{k}}\phi_{k}(\zeta_{k}) (F(\zeta_{k})+ G(\zeta_{k})\hat{u}(\zeta_{k}, \hat{W}_{c}))$, $\rho_{k}\coloneqq 1+\gamma_{c}\omega_{k}^{\top}\omega_{k}$, $\gamma_{c}\in \mathbb{R}_{>0}$ is a constant normalization gain, $\beta \in \mathbb{R}_{>0}$ is a constant forgetting factor, and $k_{c} \in \mathbb{R}_{>0}$ is a constant adaptation gain. Note that the update laws in \eqref{eq:weightsUpdate} rely on randomly sampled off-policy states $\zeta_{k}$, not the actual state $\zeta$. To guarantee the safety of the closed-loop system in \eqref{eq:dynamics_x}, the control signal is then designed as
\begin{equation}\label{eq:uControl}
    u(t) = \hat{u}(\hat{\zeta}(t),\hat{W}_{c}(t)), \quad \forall t \geq t_{0}.
\end{equation}
The following section analyzes the convergence properties of the closed-loop system and establishes local uniform ultimate boundedness of its trajectories.

\section{Stability Analysis}\label{section:stabilityAnalysis}
In this section, stability analysis of the safe observer-based SLOC architecture is carried out using Lyapunov methods. To facilitate the stability analysis, the following rank condition is necessary.
\begin{assum}
    \label{ass:CLBCADPLearnCond}There exists a constant $\underline{c}_{1}$ such that the finite set of trajectories $\left\{\zeta_{k}: \mathbb{R}_{\geq 0} \to \Omega \mid k=1,\hdots,N\right\}$ satisfies
    \begin{equation}
    0 < \underline{c}_{1} \leq\inf_{t\in[0, \infty)}\lambda_{\min}\left(\frac{1}{N}\sum_{k=1}^{N}\frac{\omega_{k}\left(t\right)\omega_{k}^{\top} \left(t\right)}{\rho_{k}^{2}\left(t\right)}\right).\label{eq:CLBCPE2}
    \end{equation}
\end{assum}
As described in \cite{SCC.Mahmud.Nivison.ea2021}, since $\omega_{k}$ is a function of $\zeta_{k}$ and $\hat{W}_{c}$,  Assumption \ref{ass:CLBCADPLearnCond} cannot be guaranteed a priori. However, unlike the PE condition utilized in results such as \cite{SCC.Vamvoudakis.Lewis2010}, Assumption~\ref{ass:CLBCADPLearnCond} can be verified online. Furthermore, since $\lambda_{\min}\left(\sum_{k=1}^{N}\frac{\omega_{k}\left(t\right)\omega_{k}^{\top} \left(t\right)}{\rho_{k}^{2}\left(t\right)}\right)$ is non-decreasing in the number of samples, $N$, Assumption \ref{ass:CLBCADPLearnCond} can be met, heuristically, by increasing the number of 
extrapolation trajectories. The calculation of a precise bound on the number of extrapolation trajectories required is outside the scope of this paper. 

The following lemma presents an alternate form of the BE to facilitate the stability analysis. For notational brevity, let $\hat{F} \coloneqq  F(\hat{\zeta})$, $\hat{G} \coloneqq  G(\hat{\zeta})$, $\nabla_{\hat{\zeta}}\phi \coloneqq \nabla_{\hat{\zeta}}\phi(\hat{\zeta})$, $\nabla_{\hat{\zeta}}B_{r} \coloneqq \nabla_{\hat{\zeta}}B_{r}(\hat{\zeta})$, and $\nabla_{\hat{\zeta}}\epsilon \coloneqq \nabla_{\hat{\zeta}}\epsilon(\hat{\zeta})$. Furthermore, the dependence on $\hat{\zeta}$, $\zeta_{k}$, $\hat{W}_{c}$, and $t$ is omitted henceforth whenever it is clear from context.
\begin{lem}\label{lem:unmeasurableBE}
     Given the critic NN weight estimation error $\tilde{W}_{c} \coloneqq W -\hat{W}_{c}$, the BE can alternatively be expressed in its unmeasurable form as
\begin{equation}\label{eq:BEAlternate}
    \hat{\delta} \coloneqq -\omega^{\top} \tilde{W}_{c} + \Delta,
\end{equation}
where $\omega \coloneqq \nabla_{\hat{\zeta}}\phi (\hat{F}+ \hat{G}\hat{u}(\hat{\zeta},\hat{W}_{c}))$, $\Delta \coloneqq -\nabla_{\hat{\zeta}}\epsilon(\hat{F} + \hat{G}u^{*}(\hat{\zeta})) + \nabla_{\hat{\zeta}}B_{r}(\hat{F}+\hat{G}\hat{u}(\hat{\zeta},\hat{W}_{c})) + \overline{u}W^{\top} \nabla_{\hat{\zeta}}\phi\hat{G}(\tanh(D^{*}(\hat{\zeta}))- \tanh(\hat{D}(\hat{\zeta},\hat{W}_{c})))+2\overline{u}^{2}\overline{R}(c_{D}^{*}-\hat{c}_{D}) + \overline{u}^{2}\overline{R}(\hat{\epsilon}_{D}-\epsilon_{D}^{*})$, $c_{D}^{*}$ and $\hat{c}_{D}$ are bounded approximation errors defined as $c_{D}^{*} \coloneqq D^{*}(\hat{\zeta})\sign(D^{*}(\hat{\zeta}))-D^{*}(\hat{\zeta})\tanh(D^{*}(\hat{\zeta}))$ and $\hat{c}_{D} \coloneqq \hat{D}\hat{\zeta},\hat{W}_{c})\sign(\hat{D}(\hat{\zeta},\hat{W}_{c}))-\hat{D}(\hat{\zeta},\hat{W}_{c})\tanh(\hat{D}(\hat{\zeta},\hat{W}_{c}))$, respectively, and $\epsilon_{D}^{*}$ and $\hat{\epsilon}_{D}$ are errors obtained when approximating the $\sign(\cdot)$ function with the $\tanh(\cdot)$ function, which satisfy $\|\epsilon_{D}^{*}\| \leq \ln(4)$ and $\|\hat{\epsilon}_{D}\| \leq \ln(4)$ so that $\ln(\boldsymbol{1}-\tanh^{2}(D^{*}(\hat{\zeta}))) = \ln(4)-2D^{*}(\hat{\zeta})\sign(D^{*}(\hat{\zeta}))+\epsilon_{D}^{*}$ and \ $\ln(\boldsymbol{1}-\tanh^{2}(\hat{D}(\hat{\zeta},\hat{W}_{c}))) = \ln(4)-2\hat{D}(\hat{\zeta},\hat{W}_{c})\sign(\hat{D}(\hat{\zeta},\hat{W}_{c}))+\hat{\epsilon}_{D}$, respectively.
\end{lem}
\begin{pf}
Substituting $\hat{\zeta}$ for $\zeta$ in \eqref{eq:optimalV} and taking its gradient with respect to $\hat{\zeta}$, followed by taking the gradient of \eqref{eq:VAppp} with respect to $\hat{\zeta}$, then substituting $\hat{\zeta}$ for $\zeta$ in the HJB equation in \eqref{eq:HJB} and subtracting it from \eqref{eq:BE1}, and finally substituting the gradients into the resulting expression, yields
\begingroup\medmuskip=0mu\thinmuskip=2mu\thickmuskip=2mu\begin{multline}\label{eq:BE2}
    \scalebox{0.95}{$\hat{\delta}(\hat{\zeta}, \hat{W}_{c}) = \hat{W}_{c}^{\top} \nabla_{\hat{\zeta}}\phi\bigl(\hat{F}+\hat{G}\hat{u}(\hat{\zeta}, \hat{W}_{c})\bigr) + U(\hat{u}(\hat{\zeta}, \hat{W}_{c}))$} \\\scalebox{0.95}{$+ \nabla_{\hat{\zeta}}B_{r}\bigl(\hat{F}+\hat{G}\hat{u}(\hat{\zeta}, \hat{W}_{c})\bigr) - W^{\top}\nabla_{\hat{\zeta}}\phi\bigl(\hat{F}+\hat{G}u^{*}(\hat{\zeta})\bigr)$} \\-  \scalebox{0.95}{$\nabla_{\hat{\zeta}}B_{r}\bigl(\hat{F} + \hat{G}u^{*}(\hat{\zeta})\bigr)  - \nabla_{\hat{\zeta}}\epsilon\bigl(\hat{F}+\hat{G}u^{*}(\hat{\zeta})\bigr) - U(u^{*}(\hat{\zeta})).$}
\end{multline}\endgroup
Substituting  equation \eqref{eq:optimalcontrol} and \eqref{eq:uApp} in \eqref{eq:controlIntegrandfunction} yields alternate representations of the optimal and approximate forms of the function $U(u)$, given by
\begin{multline} \label{eq:optimalstarU}
U(u^{*}(\hat{\zeta})) = {\overline{u} }\nabla_{\hat{\zeta}}{V^{*}}(\hat{\zeta}) G(\hat{\zeta})\tanh\bigl(D^{*}(\hat{\zeta})\bigr)\\+{\overline{u} }^{2}\overline{R}\ln\bigl(\boldsymbol{1}-\tanh^{2}\bigl(D^{*}(\hat{\zeta})\bigr)\bigr) \text{ and}
\end{multline}
\begin{multline}\label{eq:approxU}
U(\hat{u}(\hat{\zeta},\hat{W}_{c})) = {\overline{u} }\nabla_{\hat{\zeta}}\hat{V}(\hat{\zeta},\hat{W}_{c}) G(\hat{\zeta})\tanh\bigl(\hat{D}(\hat{\zeta},\hat{W}_{c})\bigr)\\+{\overline{u} }^{2}\overline{R}\ln\bigl(\boldsymbol{1}-\tanh^{2}\bigl(\hat{D}(\hat{\zeta},\hat{W}_{c})\bigr)\bigr),
\end{multline}
respectively, where $\overline{R} \coloneqq [r_{1},\hdots,r_m] \in \mathbb{R}^{1\times m}$ and $\boldsymbol{1} \in \mathbb{R}^{m}$ denotes a column vector having all of its elements equal to one. Subtracting \eqref{eq:optimalstarU} from \eqref{eq:approxU} yields 
\begin{multline}\label{eq:diffU}
   U(\hat{u}(\hat{\zeta},\hat{W}_{c}))-U(u^{*}(\hat{\zeta}))  = \overline{u} \hat{W}_c^{\top}\nabla_{\hat{\zeta}}\phi\hat{G}\tanh(\hat{D}(\hat{\zeta},\hat{W}_{c})) \\+ 2\overline{u}^{2}\overline{R}\big(D^{*}(\hat{\zeta})\sign(D^{*}(\hat{\zeta}))-\hat{D}(\hat{\zeta},\hat{W}_{c})\sign(\hat{D}(\hat{\zeta},\hat{W}_{c}))\big) \\+ \overline{u}\nabla_{\hat{\zeta}}B_{r}\hat{G}\tanh(\hat{D}(\hat{\zeta},\hat{W}_{c})) 
    -\overline{u}W^{\top} \nabla_{\hat{\zeta}}\phi\hat{G}\tanh(D^{*}(\hat{\zeta})) \\\qquad \qquad \ \
      - \overline{u}\nabla_{\hat{\zeta}}B_{r}\hat{G}\tanh(D^{*}) - \overline{u}\nabla_{\hat{\zeta}}\epsilon\hat{G}\tanh(D^{*}(\hat{\zeta})) \\+ \overline{u}^{2}\overline{R}\left(\hat{\epsilon}_{D}-\epsilon_{D}^{*}\right),
\end{multline} 
where $\epsilon_{D}^{*}$ and $\hat{\epsilon}_{D}$ are approximation errors introduced in \eqref{eq:BEAlternate}.
By using the approximation errors $c_{D}^{*}$ and $\hat{c}_{D}$ obtained when approximating the $\sign(\cdot)$ function with a $\tanh(\cdot)$ function introduced in \eqref{eq:BEAlternate}  (cf. \cite{SCC.Modares.Lewis.ea2013}), therefore,
\begingroup\medmuskip=0mu\begin{multline}\label{eq:diffD}
D^{*}(\hat{\zeta})\sign(D^{*}(\hat{\zeta})) - \hat{D}(\hat{\zeta},\hat{W}_{c})\sign(\hat{D}(\hat{\zeta},\hat{W}_{c}))
    = \\  \; \frac{R^{-1}}{2\overline{u}}\bigl(W^{\top}\nabla_{\hat{\zeta}}\phi + \nabla_{\hat{\zeta}}B_{r} + \nabla_{\hat{\zeta}}\epsilon\bigr)\hat{G}\tanh(D^{*}(\hat{\zeta})) + c_{D}^{*}  \\
    -\frac{R^{-1}}{2\overline{u}}\big(\hat{W}_{c}^{\top}\nabla_{\hat{\zeta}}\phi+\nabla_{\hat{\zeta}}B_{r}\big)\hat{G}\tanh(\hat{D}(\hat{\zeta},\hat{W}_{c}))  - \hat{c}_{D}.
\end{multline}\endgroup
After substituting \eqref{eq:diffD} into \eqref{eq:diffU} and then substituting \eqref{eq:VAppp}, \eqref{eq:uApp} and \eqref{eq:diffU} into \eqref{eq:BE2}, the BE can be expressed as given in \eqref{eq:BEAlternate}. 
\end{pf}
Note that, the residual term $\Delta$ in \eqref{eq:BEAlternate} is bounded for all $\hat{\zeta} \in \Omega$ and $\hat{W}_{c} \in \mathbb{R}^{L}$ as  \begingroup\medmuskip=0mu\begin{multline}\label{eq:deltaBound}
\Delta \leq  (\overline{F} + \overline{u}\overline{G})(\overline{\nabla \epsilon} +  \overline{\nabla B_{r}}) + 2\overline{W^{\top}}\overline{\nabla \phi}\overline{G} \\ + 2\overline{u}^{2}\big\|\overline{R}\big\|\left\|(c_{D}^{*}-\hat{c}_{D})\right\| + \overline{u}^{2}\big\|\overline{R}\big\|\left\|(\hat{\epsilon}_{D}-\epsilon_{D}^{*})\right\| \eqqcolon \overline{\Delta},
\end{multline}\endgroup
where $\overline{F} \in \mathbb{R}_{>0} $ and  $\overline{G} \in \mathbb{R}_{>0}$ are upper bounds satisfying $\sup_{\hat{\zeta} \in \Omega}\|F(\hat{\zeta})\| \leq \overline{F}$ and $\sup_{\hat{\zeta} \in \Omega}\|G(\hat{\zeta)}\| \leq \overline{G}$, respectively, which follows since the derivatives of the functions $f$ and $g$ are bounded from Assumption~\ref{ass:jacobianbounds} and since $B_{r}$ is continuously differentiable in $\hat{\zeta} \in \Omega$, its gradient$ \nabla_{\zeta}B_{r}$ is both continuous and bounded within the compact set $\Omega$. Hence, by the Extreme Value Theorem, there exist some constant $\overline{\nabla B_{r}} \in \mathbb{R}_{> 0}$ such that $\sup_{\hat{\zeta} \in \Omega}\|\nabla_{\hat{\zeta}}B_{r}(\hat{\zeta})\| \leq \overline{\nabla B_{r}}$. For notational brevity, let $F_{k} \coloneqq  F(\zeta_{k})$, $G_{k} \coloneqq  G(\zeta_{k})$, $u_{k}^{*} \coloneqq u^{*}(\zeta_{k})$, $\hat{u}_{k} \coloneqq \hat{u}(\zeta_{k}, \hat{W}_{c})$,  $D_{k}^{*} \coloneqq D^{*}(\zeta_{k})$, $\hat{D}_{k} \coloneqq \hat{D}(\zeta_{k}, \hat{W}_{c})$, $\nabla_{\zeta_{k}}\phi_{k} \coloneqq \nabla_{\zeta_{k}}\phi(\zeta_{k})$, $\nabla_{\zeta_{k}}B_{r_{k}} \coloneqq \nabla_{\hat{\zeta}}B_{r}(\zeta_{k})$, and $\nabla_{\zeta_{k}}\epsilon_{k} \coloneqq \nabla_{\hat{\zeta}}\epsilon(\zeta_{k})$. Following the same argument as Lemma~\ref{lem:unmeasurableBE}, by substituting \eqref{eq:VAppp} and \eqref{eq:uApp} into \eqref{eq:HJB}, and subtracting from \eqref{eq:BE1}, the extrapolated BE can be expressed in an unmeasurable form as
\begin{equation}\label{eq:deltaBE}
    \hat{\delta}_{k} \coloneqq -\omega_{k}^{\top} \tilde{W}_{c}+ \Delta_{k},
\end{equation}
where $\omega_{k}$ is introduced below \eqref{eq:weightsUpdate}, $\Delta_{k} \coloneqq -\nabla_{\zeta_{k}}\epsilon_{k}^{\top}(F_{k} + G_{k}u_{k}^{*})+ \nabla_{\zeta_{k}}B_{r_{k}}^{\top}(F_{k} + G_{k}\hat{u}_{k}))+{\overline{u}}W^{\top} \nabla_{\zeta_{k}}\phi_{k}G_{k}  (\tanh(D_{k}^{*})- \tanh(\hat{D}_{k}))+2{\overline{u} }^{2}\overline{R}(c_{D_{k}}^{*}-\hat{c}_{D_{k}}) +{\overline{u} }^{2}\overline{R}(\hat{\epsilon}_{D_{k}}-\epsilon_{D_{k}}^{*})$, and $c_{D_{k}}^{*}$, $\hat{c}_{D_{k}}$, $\hat{\epsilon}_{D_{k}}$, and $\epsilon_{D_{k}}^{*} \in \mathbb{R}$ are bounded approximation errors. Similar to \eqref{eq:deltaBound}, note that $\sup_{\hat{\zeta} \in \Omega}\|\Delta_{k}(\cdot)\| \leq \overline{\Delta}_{k}$ for all $t \geq t_{0}$, where $\overline{\Delta}_{k} \in \mathbb{R}_{>0}$ is a constant.

Let $Z \in \mathbb{R}^{2n+L}$ represent the concatenated state vector of the closed-loop system, defined as $Z \coloneqq \begin{bmatrix} x^{\top} & \tilde{x}^{\top} & {\tilde{W}_c}^{\top} \end{bmatrix}^{\top}$, and let $V_{L}: \mathbb{R}^{2n+L} \times \mathbb{R}_{\geq 0} \to \mathbb{R}$ be a continuously differentiable candidate Lyapunov function defined as
 \begin{equation}\label{eq:Lyap}
     V_{L}\left(Z,t\right) \coloneqq \mathcal{V}(Z,t) + \Theta(\tilde{W}_{c}, t),
 \end{equation}
where $\Theta: \mathbb{R}^{L} \times \mathbb{R}_{\geq 0} \to \mathbb{R}$ be defined as $\Theta(\hat{W}_{c}, t) \coloneqq \frac{1}{2}\tilde{W}_{c}^{\top} \Gamma^{-1}\left(t\right)\tilde{W}_{c}$ and $V: \mathbb{R}^{2n+L} \times \mathbb{R}_{\geq 0} \to \mathbb{R}$ is defined as $\mathcal{V}(Z, t) = V_{t}^{*}(x, t) + \tilde{x}^{\top}P\tilde{x}$. The function $V_{t}^{*}$ is the positive definite non-autonomous form of the optimal function $V^{*}$, defined as $V_{t}^{*}(x,t) \coloneqq V^{*}([x^{\top}, \xi(t)^{\top}]^{\top})$, which according to \cite[Lemma~1]{SCC.Kamalapurkar.Dinh.ea2015} serves as a valid Lyapunov function, being both positive definite and decrescent. In addition, similar to the results of \cite[Lemma~2]{SCC.Kamalapurkar.Dinh.ea2015}, the function $V_{t}$ satisfies the properties $V_{t}^*(0, t) = 0$ and $\underline{v}(\|x\|) \leq V_{t}^*(x, t) \leq \overline{v}(\|x\|)$ for all $x \in \mathbb{R}^n$ and $t \in [0, \infty)$, where $\underline{v}, \overline{v} : \mathbb{R}_{\geq 0} \to \mathbb{R}_{\geq 0}$ are $\mathcal{K}$ functions. The choice of $V_{t}^{*}$ as a candidate Lyapunov function is motivated by the unsuitability of the optimal value function due to its positive semi-definiteness.

 Since the candidate Lyapunov function is positive definite, \cite[Lemma 4.3]{SCC.Khalil2002} and the bound in \eqref{eq:OFBADP1Gammabound} can be used to conclude that it is bounded as
\begin{equation}
\underline{\upsilon}_{l}\left(\left\Vert Z\right\Vert \right)\leq V_{L}\left(Z,t\right)\leq\overline{\upsilon}_{l}\left(\left\Vert Z\right\Vert \right),\label{eq:OFBADPVBound}
\end{equation}
for all $t \geq t_{0}$ and for all $Z \in \mathbb{R}^{2n+L}$, where $\underline{\upsilon}_{l},\overline{\upsilon}_{l}: \mathbb{R}_{\geq 0}\rightarrow \mathbb{R}_{\geq 0}$ are class $\mathcal{K}$ functions. 

 The Lie derivative of $\mathcal{V}$ is given by
\begin{multline}
     \dot{\mathcal{V}}(Z, t) = \nabla_{\zeta}V^{*}\bigl(F + Gu^{*}(\zeta)\bigr) + \dot{V}_{x}\left(\tilde{x},x,u\right) \\+ \nabla_{\zeta}V^{*}G\bigl(\hat{u}(\hat{\zeta}, \hat{W}_{c})-u^{*}(\zeta)\bigr) .
 \end{multline}
 Note that, by the Cauchy-Schwarz inequality, Young's inequality, and the triangle inequality, the norm of the difference between the approximate control policy $\hat{u}$ and the optimal policy $u^{*}$ is given by,
\begin{multline}\label{eq:tanHDiff}
\big\|\hat{u}(\hat{\zeta}, \hat{W}_{c})-u^{*}(\zeta)\big\| \\ \quad =  \big\| {\overline{u}}\tanh(D^{*}(\zeta))- \overline{u}\tanh(\hat{D}(\hat{\zeta}, \hat{W}_{c}))\big\|   \\
\leq \ell_{g\phi r}\overline{W}\Big\|\tilde{x}\Big\| + \overline{G_{\phi R}}\bigl\|\tilde{W}_{c}\bigr\| + \frac{\lambda_{\max}(R^{-1})\overline{G}}{2}\overline{\nabla\epsilon},
\end{multline}
for all $\zeta, \hat{\zeta} \in \Omega$ and $\hat{W}_{c} \in \mathbb{R}^{L}$, where $\sup_{\zeta \in \Omega} G_{\phi R}(\cdot) \coloneqq \frac{R^{-1}G(\cdot)^{\top}}{2}\nabla_{(\cdot)}\phi(\cdot)^{\top} \leq \overline{G_{\phi R}}$, $\ell_{g\phi r} \in \mathbb{R}_{>0}$ is a constant bound over the set $\Omega$ such that $\|G_{\phi R}(\zeta)-G_{\phi R}(\hat{\zeta})\| \leq \ell_{g\phi r}\|\tilde{x}\|$ and $\overline{G} \in \mathbb{R}_{>0}$ is a positive bound such that $\sup_{\zeta \in \Omega}\|G(\zeta)\| \leq \overline{G}$.
Thus, provided that the basis functions $\phi$ are selected such that the weights $W$, the value function approximation error $\epsilon$, the basis itself $\phi$, and their gradients with respect to $\zeta$ and $\hat{\zeta}$ are bounded on the compact set $\Omega$ as described below \eqref{eq:optimalV}, by substituting the bounds in \eqref{eq:VeIneq} and \eqref{eq:tanHDiff}, using the fact that $\nabla_{\zeta}V^{*}(\zeta)(F(\zeta) + G(\zeta)u^{*}(\zeta)) =  - r(\zeta, u^{*}(\zeta))$ from the substitution of \eqref{eq:valuefunction} and \eqref{eq:optimalcontrol} into \eqref{eq:HJB}, and using 
 the positive definiteness of the functions $U$ and $B_{r}$, the Lie derivative of the function $\mathcal{V}$ is bounded by
\begingroup\medmuskip=0mu\thinmuskip=1mu\begin{equation}\label{eq:vaBound}
     \dot{\mathcal{V}}(Z, t) \leq  -\underline{q}\left(\|x\|\right)-\alpha\lambda_{\min}(P)\left\|\tilde{x}\right\|^{2} + \varpi_{1}\bigl\|\tilde{W}_{c}\bigr\| + \iota_{1}, 
\end{equation}\endgroup
for all $t \geq t_{0}$, $x \in \mathcal{X}$, $\tilde{x} \in \mathcal{B}(0,\chi)$, and $\tilde{W}_c \in \mathbb{R}^L$, where $\varpi_{1} \coloneqq \overline{G}\overline{W}\overline{\nabla\phi}\overline{G}_{\phi R}+ \overline{G}\overline{G}_{\phi R}\overline{\nabla\epsilon}$ and $\iota_{1} \coloneqq \frac{\overline{G}^{2}\lambda_{\max}(R^{-1})}{2}\overline{W}\overline{\nabla\phi}\overline{\nabla\epsilon} + \frac{\overline{G}^{2}\lambda_{\max}(R^{-1})}{2}\overline{\nabla\epsilon}^{2}
    +\frac{(\ell_{g\phi r}\overline{G}\overline{W}\overline{\nabla\phi}\overline{W}+\overline{G}\ell_{g\phi r}\overline{W}\overline{\nabla\epsilon})^{2}}{4\alpha\lambda_{\min}(P)}$. Note that the bound in \eqref{eq:vaBound} relies on the fact that $x \in \mathcal{X}$ and $\tilde{x} \in \mathcal{B}(0,\chi)$ implies that $\zeta,\hat{\zeta} \in \Omega$.
    
  As shown in \cite[Lemma~1]{SCC.Kamalapurkar.Rosenfeld.ea2016}, provided (\ref{ass:CLBCADPLearnCond}) holds and $\lambda_{\min}\{{\Gamma_{0}^{-1}}\}> 0$, the update laws in \eqref{eq:weightsUpdate}, ensure that the least squares update law satisfies
\begin{equation}\label{eq:OFBADP1Gammabound}\underline{\Gamma}\mathrm{I}_{L}\leq\Gamma\leq\overline{\Gamma}\mathrm{I}_{L},		\end{equation}
for all $t \geq t_{0}$ and some $\overline{\Gamma},\underline{\Gamma} \in \mathbb{R}_{>0}$, where $\mathrm{I}_{L}$ is a $L$ by $L$ identity matrix. Using the least square gain update law in \eqref{eq:weightsUpdate} and the bound in \eqref{eq:OFBADP1Gammabound}, the normalized regressor $\frac{\omega_{k}}{\rho_{k}}$ is bounded as $\sup_{t \geq t_{0}}\|\frac{\omega_{k}}{\rho_{k}}\| \leq \frac{1}{2\sqrt{\gamma_{c}\underline{\Gamma}}}$. The Lie derivative of $\Theta$ along the flow of \eqref{eq:weightsUpdate} is given by
 \begin{equation}
     \dot{\Theta}(\tilde{W}_{c}, t) \coloneqq - \tilde{W}_{c}^{\top} \Gamma^{-1}\dot{\hat{W}}_{c} -\frac{1}{2} \tilde{W}_{c}^{\top} \Gamma^{-1}\dot{\Gamma}\Gamma^{-1}\tilde{W}_{c}.
 \end{equation}
 Substituting the update laws in \eqref{eq:weightsUpdate} into the derivative and simplifying yields
 \begin{multline}
       \dot{\Theta}(\tilde{W}_{c}, t) = - \frac{k_{c}}{N}\tilde{W}_{c}^{\top}\sum_{k=1}^{N}\frac{\omega_{k}\omega_{k}}{\rho_{k}}^{\top}\tilde{W}_{c}
      + \frac{k_{c}}{N}\tilde{W}_{c}^{\top}\sum_{k=1}^{N}\frac{\omega_{k}}{\rho_{k}}\hat{\delta}_{k} \\
     -\frac{1}{2}\beta\tilde{W}_{c}^{\top}\Gamma^{-1}\tilde{W}_{c}  + \frac{k_{c}}{2N}\tilde{W}_{c}^{\top}\sum_{k=1}^{N}\frac{\omega_{k}\omega_{k}^{\top} }{\rho_{k}^{2}}\tilde{W}_{c}. 
 \end{multline}
 Provided Assumption~\ref{ass:CLBCADPLearnCond} holds and using the fact that $\frac{\omega_{k}{\omega_{k}}^{\top} }{\rho_{k}^{2}}\leq \frac{\omega_{k}{\omega_{k}}^{\top} }{\rho_{k}}$, the Lie derivative of $\Theta$ is bounded, by
 \begin{equation}\label{eq:vwBound} \dot{\Theta}(\tilde{W}_{c}, t) \leq -k_{c}\underline{c}\bigl\|\tilde{W}_{c}\bigr\|^{2}
      + \varpi_{2}\bigl\|\tilde{W}_{c}\bigr\|, 
\end{equation}
for all $t \geq t_{0}$ and $\tilde{W}_c \in \mathbb{R}^L$, where $\underline{c} \in \mathbb{R}_{>0}$ be a constant defined as $\underline{c} \coloneqq\frac{\beta}{2\overline{\Gamma} k_{c}} + \frac{\underline{c}_{1}}{2}$, $\varpi_{2} \coloneqq \frac{k_{c}\overline{\Delta}_{k}}{2\sqrt{\gamma_{c}\underline{\Gamma}}}$.

To facilitate the development of Theorem~\ref{thm:boundedAndSafe},  let $\iota \in \mathbb{R}_{> 0}$ be a constant defined as 
$\iota \coloneqq   \frac{\overline{G}^{2}\lambda_{\max}(R^{-1})}{2}\overline{\nabla\epsilon}^{2} + \frac{\overline{G}^{2}\lambda_{\max}(R^{-1})}{2}\overline{W}\overline{\nabla\phi}\overline{\nabla\epsilon}  +\frac{\left(\varpi_{1} + \varpi_{2}\right)^{2}}{2k_{c}\underline{c}} +\frac{(\ell_{g\phi r}\overline{G}\overline{W}\overline{\nabla\phi}\overline{W}+\overline{G}\ell_{g\phi r}\overline{W}\overline{\nabla\epsilon})^{2}}{4\alpha\lambda_{\min}(P)}$ and let $\upsilon_{l}: \mathbb{R}^{2n+L} \to \mathbb{R}$ be a class $\mathcal{K}$ function defined as \begin{equation}\label{eq:classKUpsilon}
     \upsilon_{l}\coloneqq \frac{\underline{q}\left(\|x\|\right)}{2}+ \frac{\alpha\lambda_{\min}(P)}{2}\left\|\tilde{x}\right\|^{2}  + \frac{k_{c}\underline{c}}{4}\bigl\|\tilde{W}_{c}\bigr\|^{2}.
 \end{equation}
 The following theorem establishes local uniform ultimate boundedness of the trajectories of the closed-loop system. 
\begin{thm}\label{thm:boundedAndSafe}
   Let $\overline{\chi} > 0$ be such that $\mathcal{B}(0, \overline{\chi})\subset\mathcal{X}\times\mathcal{B}(0,\chi)\times\mathbb{R}^{L}$. Given the system in \eqref{eq:dynamics_zeta} controlled using the controller designed in \eqref{eq:uApp}, if
\begin{enumerate}[label=H\arabic*:]
    \item\label{hyp:jacobianbounds} Assumptions \ref{ass:jacobianbounds} and \ref{ass:CLBCADPLearnCond} hold, 
    
    \item Hypothesis \ref{hyp:observerGains} of Theorem \ref{thm:stateObserver} holds,
    
    \item the state estimate \eqref{eq:dynamics_x} is obtained using the state observer in \eqref{eq:observerdynamics_x},
    
    \item the barrier-like function $B_{r}$ is selected such that it satisfies the properties highlighted above \eqref{eq:controlBarrier},
    
    \item the control gains are selected large enough based on the sufficient condition\footnote{Although $\iota$ generally increases with increasing $\zeta$, the condition in \eqref{zeta_cond} can be satisfied provided the points for BE extrapolation are selected such that $\underline{c}$, introduced and control gain, $k_{c}$ is large enough, and the basis for the value function approximation are selected such that $\overline{\epsilon}$ and $\overline{\|\nabla{\epsilon\|}}$ are sufficiently small.} 
    \begin{gather}
      {\upsilon_{l}}^{-1}\left(\iota\right) \leq  {\overline{\upsilon}_{l}}^{-1}\left(  \underline{\upsilon}_{l}(\overline{\chi})\right),\label{zeta_cond}
    \end{gather}
    \item\label{hyp:weightsUpdate} and the weights $\hat{W}_{c}$ and least squares gain matrix $\Gamma$ are updated according to the update laws in \eqref{eq:weightsUpdate},
\end{enumerate}
then the concatenated state $Z$ is locally uniformly ultimately bounded.
\end{thm}
\begin{pf}
The Lie derivative of $V_{L}$ along the flow of \eqref{eq:dynamics_x}, \eqref{eq:augError2}, and \eqref{eq:weightsUpdate} is given by
\begin{equation}\label{eq:LyapD}
     \dot{V}_{L}\left(Z,t\right) =  \dot{\mathcal{V}}(Z, t) + \dot{\Theta}(\tilde{W}_{c}, t) 
\end{equation}
By substituting \eqref{eq:VeIneq}, \eqref{eq:vaBound} and \eqref{eq:vwBound} into \eqref{eq:LyapD} and
 applying completion of squares, the Lie derivative in \eqref{eq:LyapD} can be bounded as
\begingroup\medmuskip=0mu\thinmuskip=0mu\thickmuskip=1mu\begin{equation}\label{eq:VorbitDeriv}
     \dot{V}_{L}(Z,t) \leq-\underline{q}\left(\|x\|\right)-\alpha\lambda_{\min}(P)\left\|\tilde{x}\right\|^{2}  -\frac{k_{c}\underline{c}}{2}\bigl\|\tilde{W}_{c}\bigr\|^{2} + \iota,
 \end{equation}\endgroup
 for all $t \geq t_{0}$, $x \in \mathcal{X}$, $\tilde{x} \in \mathcal{B}(0,\chi)$, and $\tilde{W}_c \in \mathbb{R}^L$. Using \eqref{eq:classKUpsilon}, the Lie derivative can be bounded as 
\begin{equation}
     \dot{V}_{L}(Z,t) \leq -\upsilon_{l}\left(\|Z\|\right),
 \end{equation}
 for all $Z$ such that $Z \in \mathcal{B}(0, \overline{\chi})$ and $\|Z\| \geq {\upsilon_{l}}^{-1}(\iota)$.  Using the sufficient conditions stated in \eqref{zeta_cond} and the bound in \eqref{eq:OFBADPVBound}, \cite[Theorem~4.18]{SCC.Khalil2002} can be invoked to conclude that $Z$ is locally uniformly ultimately bounded. In particular, all trajectories starting from initial conditions that satisfy $\|Z(t_{0})\| \leq  {\overline{\upsilon}}^{-1}\left(  \underline{{\upsilon_{l}}}(\overline{\chi})\right)$ remain in $\mathcal{B}(0, \overline{\chi})$ for all $t \geq t_{0}$ and satisfy 
$\lim \sup_{t\to\infty} \|Z\left(t\right)\| \leq  {\overline{\upsilon}}^{-1}\left(  \underline{{\upsilon_{l}}}\left(\iota\right)\right)$. Therefore, provided $\|Z(t_{0})\| \leq  {\overline{\upsilon}}^{-1}\left(  \underline{{\upsilon_{l}}}(\overline{\chi})\right)$, the state and the state estimates, under the controller in \eqref{eq:uApp} and the observer in \eqref{eq:observerdynamics_x}, remain within the compact ball $\mathcal{B}(0, \overline{\chi})$.
\end{pf}

\section{Simulation Results}\label{section:simulation}

In this section, a simulation study is performed to demonstrate the effectiveness of the developed method using a control-affine nonlinear system of the form \eqref{eq:dynamics_x} with state $x = [(x)_{1}, (x)_{2}]^{\top}$, where
\begingroup\medmuskip=0mu\begin{equation}\label{eq:simDyn}
f(x) =\begin{bmatrix}
-(x)_{1} + (x)_{2} \\
 -\frac{1}{2}(x)_{1}-\frac{1}{2}(x)_{2}(1-(\cos(2(x)_{1})+2)^{2})
\end{bmatrix},
\end{equation}\endgroup
$g(x) = [0, \cos\left(2(x)_{1}\right)+2]$, and $C = [0, 1]$. In this study, the system state is not available for measurement, but the output is measurable. The control objective is to minimize the infinite horizon cost in \eqref{eq:costFunctional} and to drive the trajectories of the nonlinear system in \eqref{eq:simDyn} to the origin while ensuring safety. To demonstrate the efficacy of the developed method and to show the robustness of the controller to errors due to the lack of full state measurement, two simulation studies using the dynamics in \eqref{eq:simDyn} are performed.
\subsection{Ensuring safety by staying within a given set}\label{subsection:simStudy1}
The first study shows that the developed method can ensure that trajectories of the system remain within a given safe set $\mathcal{S}$ as defined in \eqref{eq:safeSet1}--\eqref{eq:safeSet3} where $h(x) = -(x)_{2}^{2} - (x)_{1} + 1$  (cf. \cite{SCC.Jankovic.ea2018}).
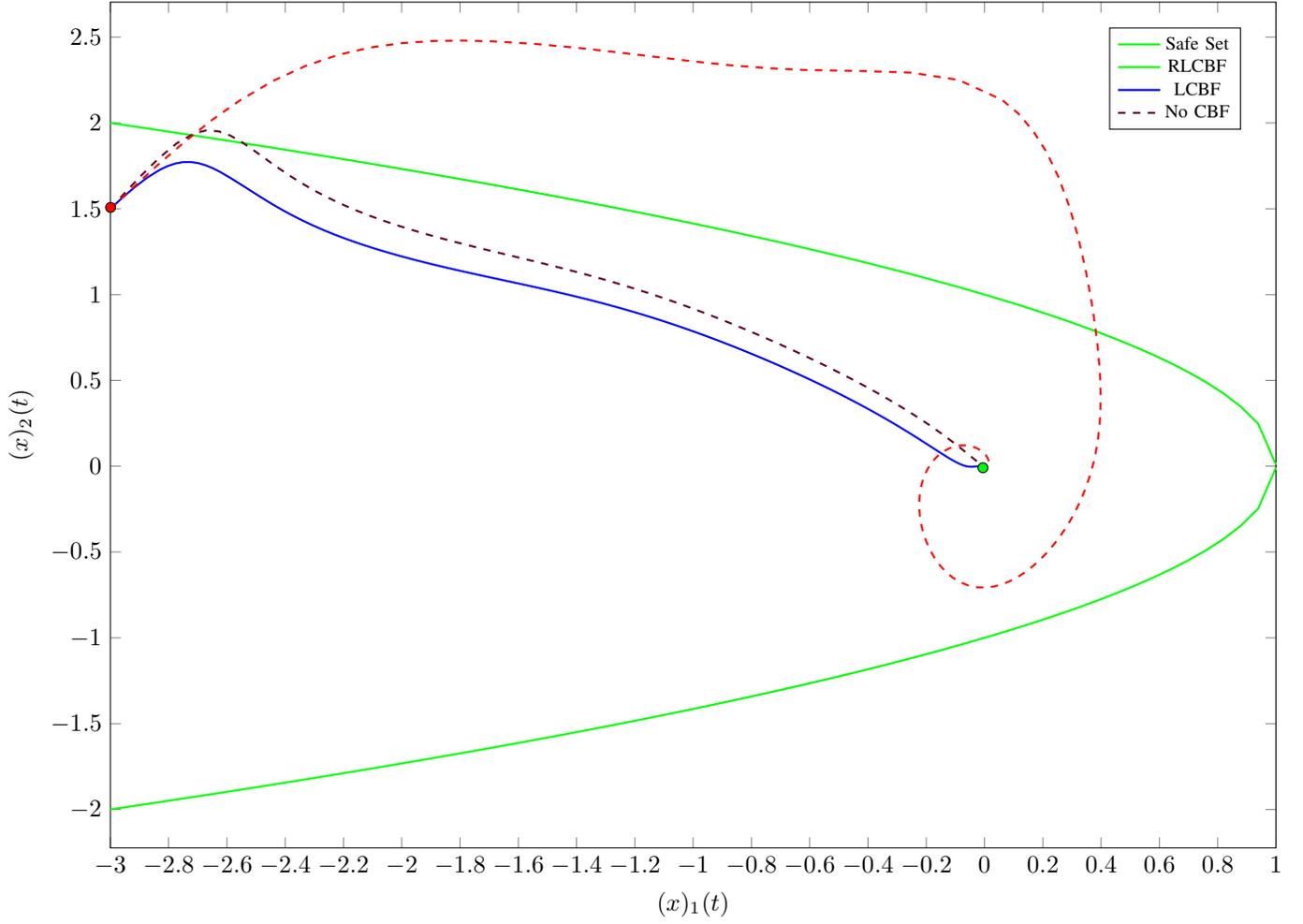
\begin{figure}
        \centering
        \input{figures/safeSetPlotSim1}
	\caption{Ensuring safety by staying within a given set (enclosed by a thick green boundary).}
		\label{fig:safetySim1}
\end{figure}
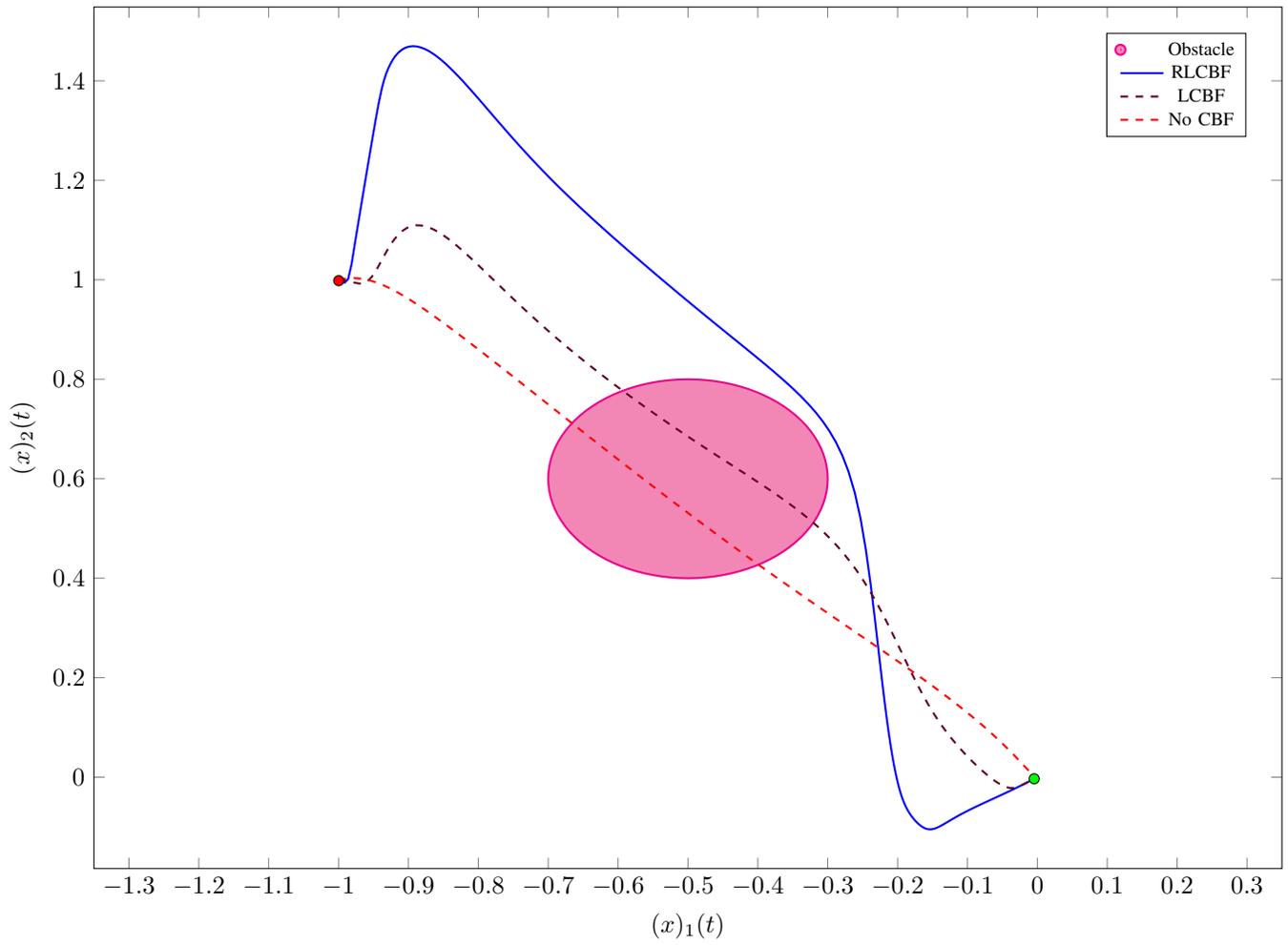
\begin{figure}
        \centering
        \input{figures/obstacleAvoidancePlotSim2}
	\caption{Obstacle avoidance.}
		\label{fig:barrierSim2}
\end{figure}
The basis for value function approximation is selected as a polynomial basis given by\footnote{The selection of a polynomial basis is motivated by the Weierstrass Approximation Theorem (see \cite[Theorem 1.5]{SCC.Sauvigny2012}) and the specific monomials used here are obtained using trial and error.}  $\phi(\hat{\zeta}) \coloneqq
[(\hat{\zeta})_{1}^{2},(\hat{\zeta})_{1}(\hat{\zeta})_{2}, (\hat{\zeta})_{2}^{2}, (\hat{\zeta})_{1}(\hat{\zeta})_{3}, (\hat{\zeta})_{1}(\hat{\zeta})_{3}, (\hat{\zeta})_{3}^{2}]^{\top}$. Initial conditions of the state and the estimated state for the first study are selected as $x(0) = [-3, 1.5]^{\top}$ and $\hat{x}(0) = [-1.5, 1]^{\top}$, respectively, with the state and the control penalty in \eqref{eq:costFunction} selected as $Q=\mathrm{I}_{2}$ and $R = 1$, respectively. The Jacobian bounds for the functions $f$ and $g$ are selected by assuming that $(x)_{1} \in [-3, 3]$ and $(x)_{2} \in [-3, 3]$ throughout the simulation, and the barrier function ensures that the bounds remain valid. To obtain the symmetric positive definite matrix, $P$, and the three observer gains, $L_{1}$, $L_{2}$, and $L_{3}$ that satisfy the stability conditions developed in Section~\ref{section:stateEstimator}, the LMI in \eqref{eq:lmi} is solved using \texttt{Sedumi} in \texttt{YALMIP} \cite{SCC.YALMIP.2004} with the learning rate $\alpha = 2$ to get the matrices $P = [0.27222, 0.15875; 0.15875, 0.40954]$, $L_{3} = [-8.82113, 11.5823]^{\top}$, $L_{1} = [0.14719, 0.14719]^{\top}$, and $L_{2} = [0.045396, 0.045396]^{\top}$. The gain for the robust barrier function in \eqref{eq:controlBarrier} is selected as $\kappa = 0.01$. Since the true values of the value function weights are unknown, the initial estimate of the weights is selected as $\hat{W}_{c}(0) = [0.5, 1, 0.8,  0.1, 0.1, 0.1]^{\top}$ with the initial least squares gain matrix selected as $\Gamma(0) = \mathrm{I}_{6}$. The saturation constraint on the control input is considered as ${\overline{u}} = 10$, the LMI parameter is selected a $\theta = \mathrm{I}_{2}$, and the learning gains are selected as $k_{c} = 5$, $\nu=0.7$, $\beta = 0.01$, $\kappa = 0.1$, $\ell = 0.1$, and $e_{0} = 2.5$.  The simulation uses 100 fixed Bellman error extrapolation points selected from a $1\times 1$  square centered around the origin of the system.  
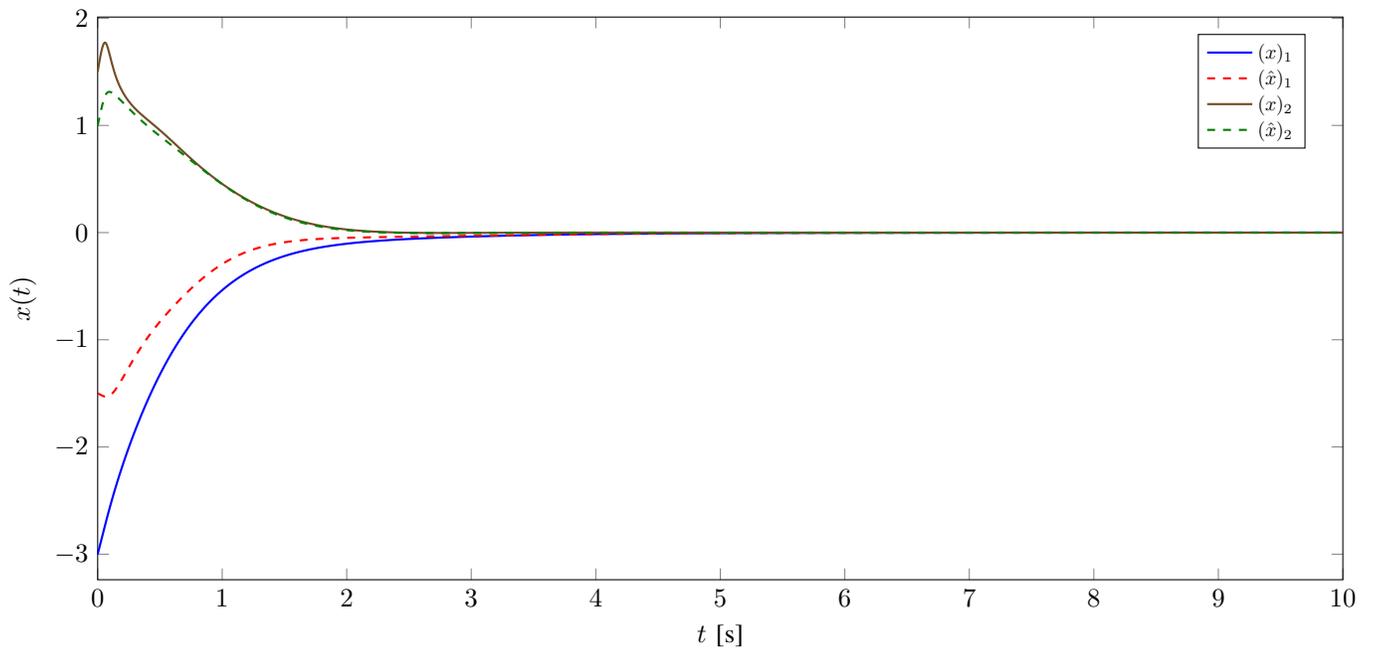
\begin{figure}
    \centering
        \centering
        \input{figures/statePlotSim1}
	\caption{The trajectories of the actual state $x$ and estimated state  $\hat{x}$ for the ensuring safety within a given set.}
		\label{fig:stateSim1}
\end{figure}
\begin{figure}
        \centering
       \input{figures/criticWeightsPlotSim1}
		\caption{The trajectories of the estimated critic NN weights for ensuring safety within a given set.}
		\label{fig:weightsSim1}
\end{figure}
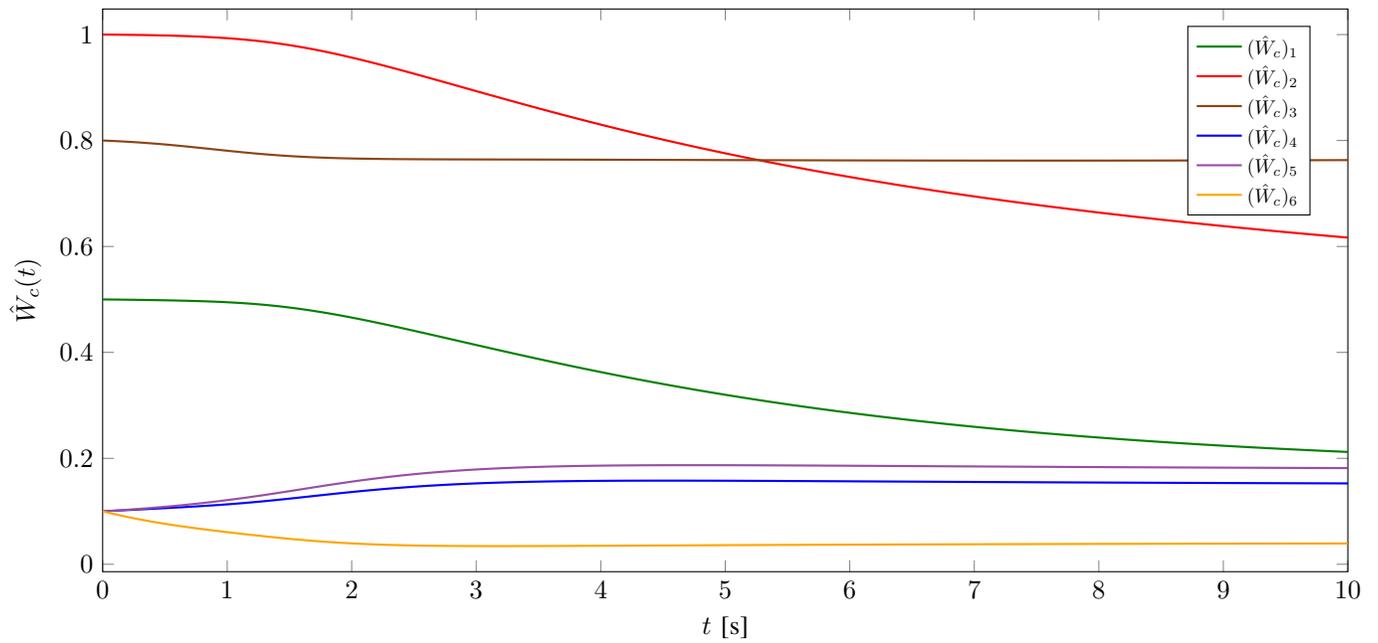
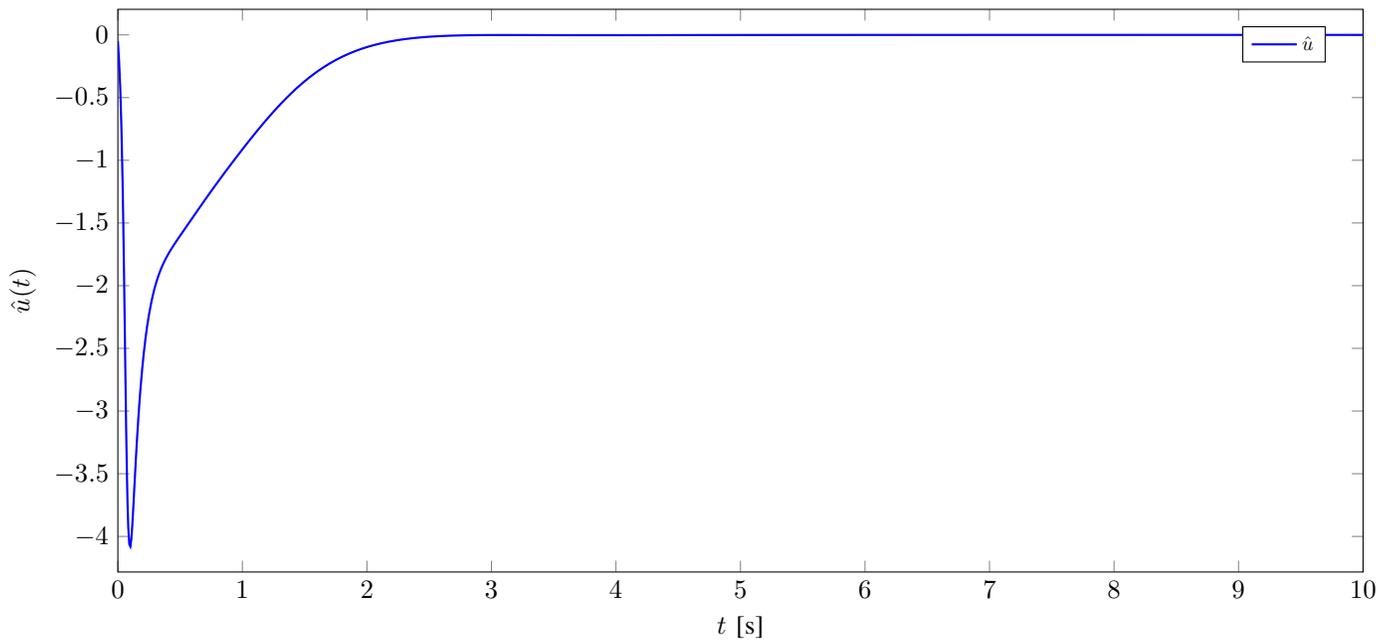
\begin{figure}
        \centering
        \input{figures/controlPlotSim1}
		\caption{The trajectory of constrained control input for ensuring safety within a given set.}
		\label{fig:controlSim1}
\end{figure}

\subsection{Obstacle Avoidance}\label{subsection:simStudy2}
For the second study, an obstacle avoidance problem is considered \eqref{eq:controlBarrier}, where the nonlinear system in \eqref{eq:simDyn} is tasked with the safety objective of avoiding a circular obstacle while being regulated to the origin using the approximate controller defined in \eqref{eq:uControl}, where the safe set in \eqref{eq:safeSet1}--\eqref{eq:safeSet3} is defined using   
    $h(x) \coloneqq \|x - z\|^{2} - r^{2}$, 
 where $z = [(z)_{1}, (z)_{2}]^{\top}$ represents the center of the circular obstacle and $r \in \mathbb{R}_{>0}$ is the radius of the set.

Initial conditions of the state and the estimated state are selected as $x(0) = [-1,1]^{\top}$ and $\hat{x}(0) = [-1.5,1.5]^{\top}$ with the state and the control penalty in \eqref{eq:costFunction} selected as $Q = \mathrm{I}_{2}$ and $R = 1$, respectively. The gain for the robust barrier function in \eqref{eq:controlBarrier} is selected as $\kappa = 2.5$. The Jacobian bounds for the functions $f$ and $g$ are selected by assuming that $(x)_{1} \in [-2, 2]$ and $(x)_{2} \in [-2, 2]$ throughout the simulation. Using this assumption, the values $P$, $L_{1}$, $L_{2}$, and $L_{3}$ are obtained as $P = 0.47897, 1.0306; 1.0306, 2.6555]$, $L_{3} = [-99.6211, 41.064]^{\top}$, $L_{1} = [0.3956, 0.13187]^{\top}$, and $L_{2} = [0.15735, 0.15735]^{\top}$, respectively. The center of the circular obstacle is selected to be $z = [-0.5, 0.6]^{\top}$ and the radius of the obstacle is selected as $r = 0.2$. The simulation uses 100 fixed Bellman error extrapolation points selected from a $2\times 2$  square centered around the origin of the system. The simulation parameters are $\ell = 0.175$, $\epsilon_{0} = 0.7$, and the rest of the initial conditions, parameters, and gains for the second simulation study are selected to be the same as the first study in Section~\ref{subsection:simStudy1}.
\begin{figure}
    \centering
        \centering
        \input{figures/statePlotSim2}
	\caption{The trajectories of the actual state $x$ and estimated state  $\hat{x}$ for the obstacle avoidance problem.}
		\label{fig:stateSim2}
\end{figure}
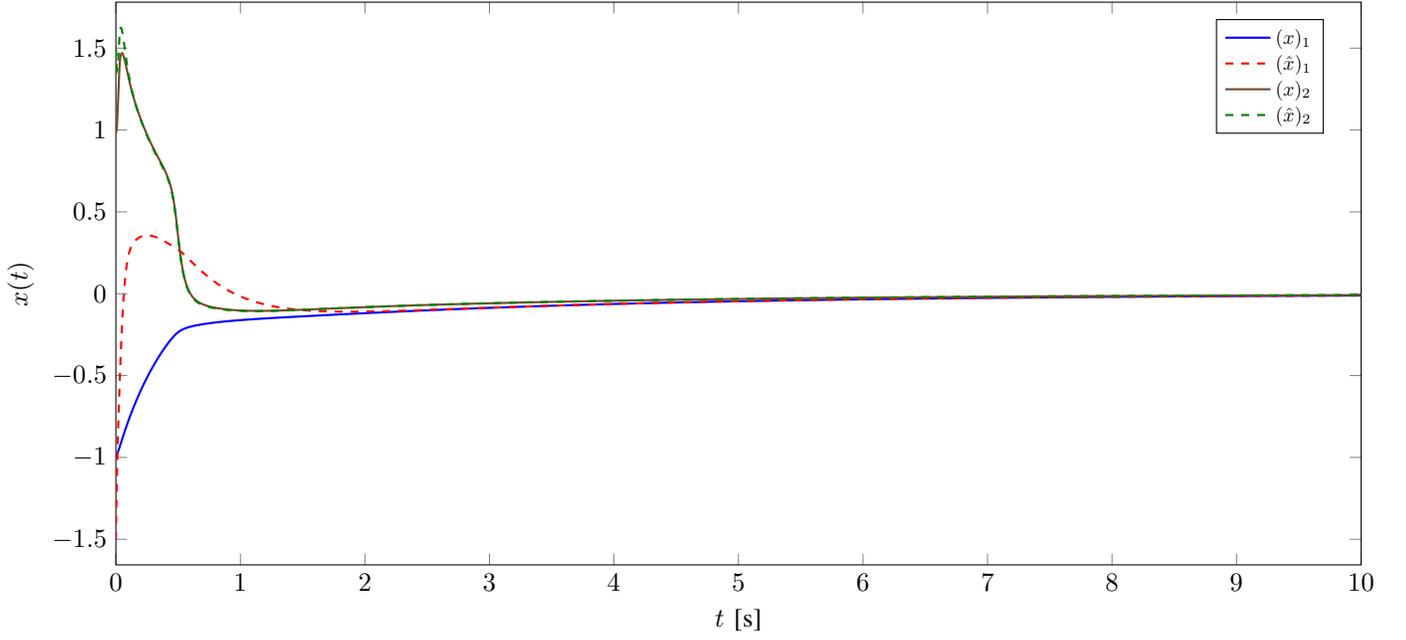
\begin{figure}
        \centering
       \input{figures/criticweightsPlotSim2}
		\caption{The trajectories of the estimated critic NN weights for the obstacle avoidance problem.}
		\label{fig:weightsSim2}
\end{figure}
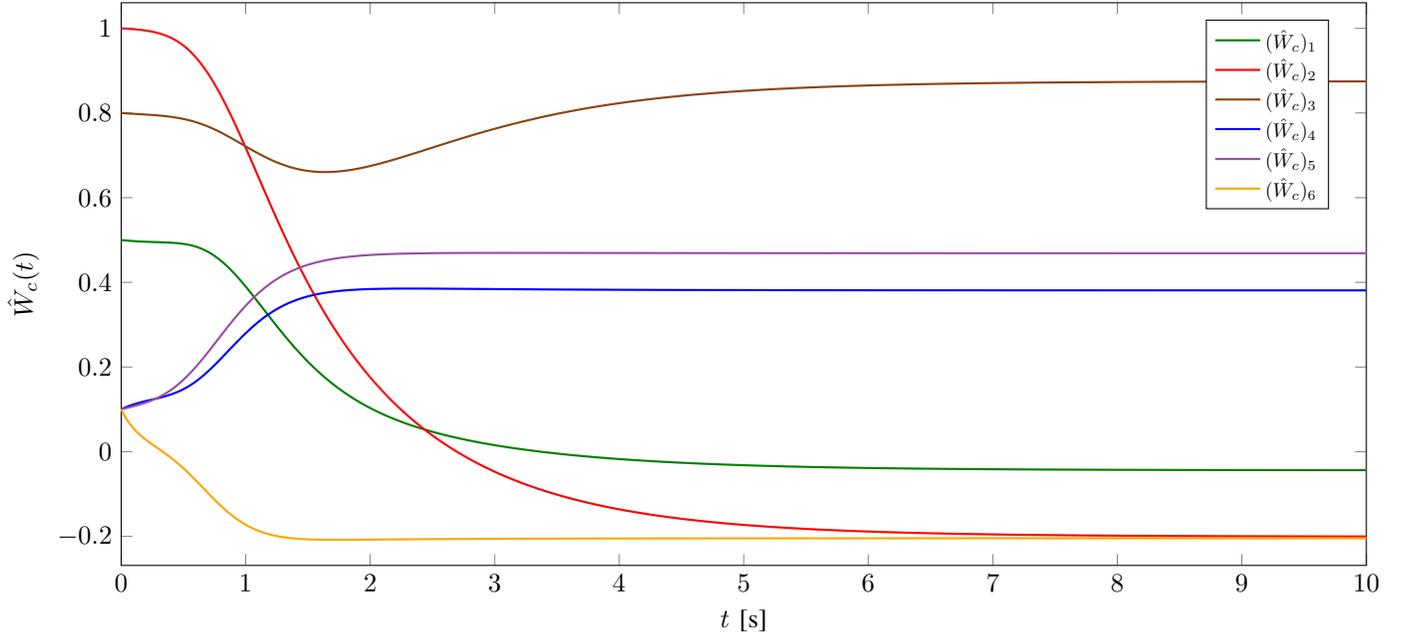
\begin{figure}
        \centering
        \input{figures/controlPlotSim2}
		\caption{The trajectory of constrained control input for the obstacle avoidance problem.}
		\label{fig:controlSim2}
\end{figure}
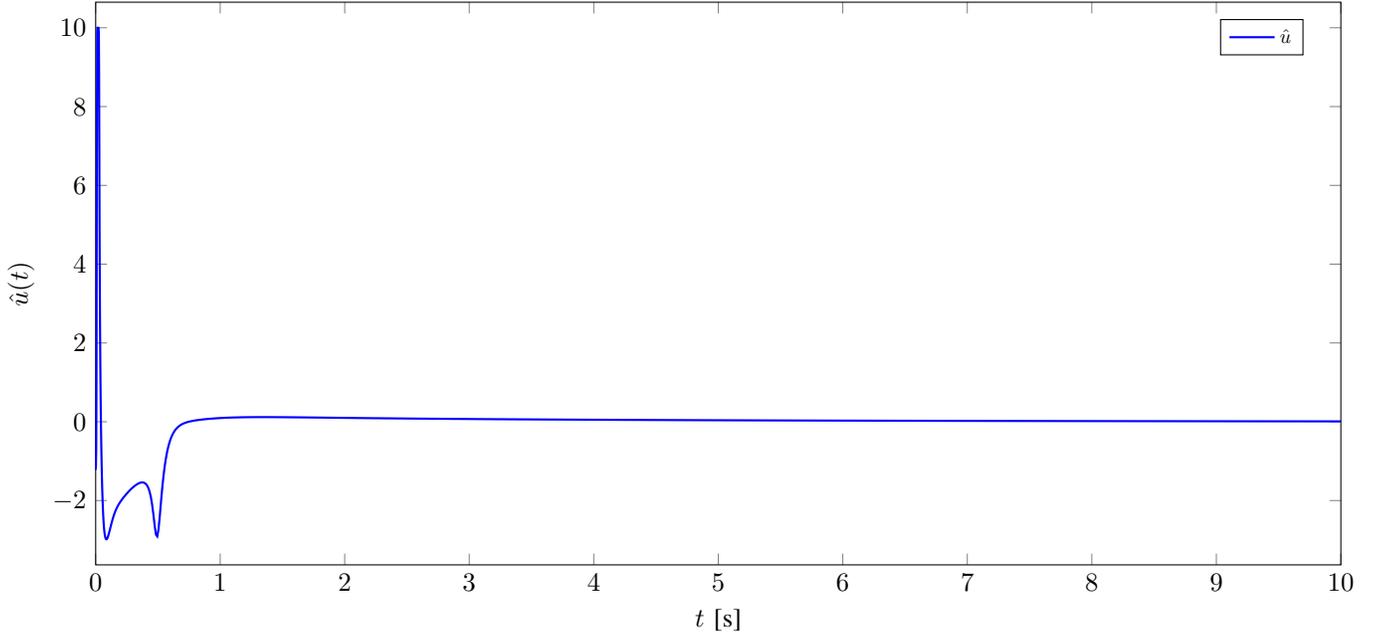

\subsection{Results} \label{ ResultSim1}
   For the first study in Section~\ref{subsection:simStudy1}, focused on ensuring safety within the set $\mathcal{S}$, it is evident from Figure~\ref{fig:safetySim1} that the RLCBF-based controller, developed in this study, adheres to the safety constraints even when utilizing the approximate controller designed in \eqref{eq:uApp}. This robust performance stems from adding the robustifying term in \eqref{eq:compensationTerm} to account for any difference between the exact and approximate feedback control policy. In contrast, relying solely on LCBFs, as depicted in Figure~\ref{fig:safetySim1}, proves inadequate in guaranteeing safety. For the obstacle avoidance problem,  it can be observed that the LCBF approach also falls short in circumventing the red circular obstacle. Conversely, the RLCBF-based method ensures that system trajectories in \eqref{eq:simDyn} steer clear of the red obstacle while regulating towards the origin. 

Figures \ref{fig:stateSim1} and \ref{fig:stateSim2} show that the trajectories of the actual state and state estimates converge to the same values and are regulated to the origin for both problems, thereby demonstrating the effectiveness of the designed LMI-based observer for state estimation of output-feedback nonlinear systems and validating the stability results developed in Section~\ref{section:stateEstimator}. 
The weight estimates, as depicted in Figures \ref{fig:weightsSim1} and \ref{fig:weightsSim2}, are shown to be UUB. While Figures \ref{fig:weightsSim1} and \ref{fig:weightsSim2} show that the learning algorithm converges quickly, it cannot be claimed that the estimates of the weights converge to their true values since the true value function and hence the exact critic weights are unknown. Figures~\ref{fig:controlSim1} and \ref{fig:controlSim2} show that the trajectory of the approximate safety-aware controller remains bounded while maintaining the stability of the closed-loop system as it simultaneously learns the optimal value function.

\section{Conclusion}\label{section:conclusion}
    In this paper, an observer-based SLOC framework for input-constrained nonlinear systems is developed using an RLCBF-based penalty introduced in the cost function of the optimal control problem. LMIs are formulated to obtain observer gain matrices, and an observer-based safe adaptive near-optimal controller is developed that guarantees safety and maintains stability during the learning phase as it seeks to find an approximate solution to the optimal control problem. 

    While the simulation results demonstrate the efficacy of our method and validate the results from the Lyapunov-based stability analysis, certain limitations to the developed method exist. The computational complexity introduced by RCBFs, sensitivity to model uncertainties, numerical issues that occur if the LMI is poorly conditioned, and the inherently conservative nature of safe controllers that utilize barrier functions to enforce safety constraints are ongoing areas of investigation. Future research will involve exploring techniques involving tunable CBFs like those in \cite{SCC.Parwana.Mustafa.ea2022} to improve system safety while being sufficiently far from the boundary of
    the safe set. Additionally, the integration of current adaptive techniques to address model uncertainties holds promise for advancing the robustness of the proposed framework. 

 \bibliographystyle{plain}

\bibliography{scc,sccmaster,scctemp}

\end{document}

%% file: figures/Cohen2020.tex
\begin{tikzpicture}
    \begin{axis}[
        axis equal,
        xlabel={$(x)_{1}(t)$},
        ylabel={$(x)_{2}(t)$},
        legend pos = north east,
        legend style={nodes={scale=0.5, transform shape}},
        enlarge y limits=0.05,
        enlarge x limits=0,
        xmin=-1.5, 
        xmax=0.1,
        width=\linewidth,
        height=0.7\linewidth,
        label style={font=\scriptsize},
        tick label style={font=\scriptsize}
    ]

        \addplot [thick, color=magenta, fill=magenta!60, circular legend] table {data/x0DataCohenWithoutCBF.dat};
        \addlegendentry{Obstacle};

        \newcommand{\plotTrajectory}[3]{%
            \addplot [#2] table {#1};
            \addlegendentry{#3}
        }

        \newcommand{\plotMarkersOnly}[2]{%
            \pgfplotstableread{#1}\datatable
            \pgfplotstablegetelem{0}{[index]0}\of{\datatable}
            \pgfmathsetmacro{\startx}{\pgfplotsretval}
            \pgfplotstablegetelem{0}{[index]1}\of{\datatable}
            \pgfmathsetmacro{\starty}{\pgfplotsretval}

            \pgfplotstablegetrowsof{\datatable}
            \pgfmathtruncatemacro{\lastindex}{\pgfplotsretval-1}
            \pgfplotstablegetelem{\lastindex}{[index]0}\of{\datatable}
            \pgfmathsetmacro{\endx}{\pgfplotsretval}
            \pgfplotstablegetelem{\lastindex}{[index]1}\of{\datatable}
            \pgfmathsetmacro{\endy}{\pgfplotsretval}

            \addplot[mark=*, mark size=2pt, color=black, fill=#2, forget plot] 
                coordinates {(\startx,\starty)};
            \addplot[mark=*, mark size=2pt, color=black, fill=green, forget plot] 
                coordinates {(\endx,\endy)};
        }

        \plotTrajectory{data/xDataCohenWithoutCBF.dat}{color=red}{$x(t)$ (No CBF)}
        \plotTrajectory{data/xHatDataCohenWithoutCBF.dat}{dashed, color=cyan}{$\hat{x}(t)$ (No CBF)}
        \plotTrajectory{data/xDataCohenWithCBF.dat}{color=purple!50!black}{$x(t)$ (CBF)}
        \plotTrajectory{data/xHatDataCohenWithCBF.dat}{dashed, color=blue}{$\hat{x}(t)$ (CBF)}

        \plotMarkersOnly{data/xDataCohenWithoutCBF.dat}{red}
        \plotMarkersOnly{data/xHatDataCohenWithCBF.dat}{blue}

    \end{axis}
\end{tikzpicture}

%% file: figures/safeSetPlotSim1.tex
\begin{tikzpicture} 
\begin{axis}[
    xlabel={$(x)_{1}(t)$},
    ylabel={$(x)_{2}(t)$},
    legend pos = north east,
    legend style={nodes={scale=0.5, transform shape}},
    enlarge y limits=0.05,
    enlarge x limits=0,
    width=\linewidth,
    height=0.7\linewidth,
    view={0}{90},
]

\addplot[
    green,
    very thick,
    domain=-2:2,
    samples=200
]
({1-x^2},{x});
\addlegendentry{Safe Set}

    \newcommand{\plotMarkersOnly}[2]{%
        \pgfplotstableread{#1}\datatable
        \pgfplotstablegetelem{0}{[index]0}\of{\datatable}
        \pgfmathsetmacro{\startx}{\pgfplotsretval}
        \pgfplotstablegetelem{0}{[index]1}\of{\datatable}
        \pgfmathsetmacro{\starty}{\pgfplotsretval}

        \pgfplotstablegetrowsof{\datatable}
        \pgfmathtruncatemacro{\lastindex}{\pgfplotsretval-1}
        \pgfplotstablegetelem{\lastindex}{[index]0}\of{\datatable}
        \pgfmathsetmacro{\endx}{\pgfplotsretval}
        \pgfplotstablegetelem{\lastindex}{[index]1}\of{\datatable}
        \pgfmathsetmacro{\endy}{\pgfplotsretval}

        \addplot[only marks, mark=*, mark size=2pt, color=black, fill=#2, forget plot]
            coordinates {(\startx,\starty)};
        \addplot[only marks, mark=*, mark size=2pt, color=black, fill=green, forget plot]
            coordinates {(\endx,\endy)};
    }

    \newcommand{\plotTrajectory}[3]{%
        \addplot [#2, mark=none] table {#1};
        \addlegendentry{#3}
    }

    \plotTrajectory{data/xDataRCBFSim1.dat}{color=blue}{$x(t)$ (RLCBF)}
    \plotTrajectory{data/xHatDataRCBFSim1.dat}{dashed, color=cyan}{$\hat{x}(t)$ (RLCBF)}
    \plotTrajectory{data/xDataCBFSim1.dat}{color=purple!50!black}{$x(t)$ (LCBF)}
    \plotTrajectory{data/xHatDataCBFSim1.dat}{dashed, color=orange!85!black }{$\hat{x}(t)$ (LCBF)}
    \plotTrajectory{data/xDataNoCBFSim1.dat}{color=red}{$x(t)$ (No CBF)}
    \plotTrajectory{data/xHatDataNoCBFSim1.dat}{dashed, color=magenta}{$\hat{x}(t)$ (No CBF)}

    \plotMarkersOnly{data/xDataRCBFSim1.dat}{blue}
    \plotMarkersOnly{data/xHatDataNoCBFSim1.dat}{magenta}
    \plotMarkersOnly{data/xDataNoCBFSim1.dat}{red}

\end{axis}
\end{tikzpicture}

%% file: figures/obstacleAvoidancePlotSim2.tex
\begin{tikzpicture}
    \begin{axis}[
        axis equal,
        xlabel={$(x)_{1}(t)$},
        ylabel={$(x)_{2}(t)$},
        legend pos = north east,
        legend style={nodes={scale=0.5, transform shape}},
        enlarge y limits=0.05,
        enlarge x limits=0.35,
        width=\linewidth,
        height=0.7\linewidth,
    ]

            \newcommand{\plotTrajectory}[3]{%
                \addplot [#2] table {#1};
                \addlegendentry{#3}
            }
    
            \newcommand{\plotMarkersOnly}[2]{%
                \pgfplotstableread{#1}\datatable
                \pgfplotstablegetelem{0}{[index]0}\of{\datatable}
                \pgfmathsetmacro{\startx}{\pgfplotsretval}
                \pgfplotstablegetelem{0}{[index]1}\of{\datatable}
                \pgfmathsetmacro{\starty}{\pgfplotsretval}
    
                \pgfplotstablegetrowsof{\datatable}
                \pgfmathtruncatemacro{\lastindex}{\pgfplotsretval-1}
                \pgfplotstablegetelem{\lastindex}{[index]0}\of{\datatable}
                \pgfmathsetmacro{\endx}{\pgfplotsretval}
                \pgfplotstablegetelem{\lastindex}{[index]1}\of{\datatable}
                \pgfmathsetmacro{\endy}{\pgfplotsretval}
    
                \addplot[mark=*, mark size=2pt, color=black, fill=#2, forget plot] 
                    coordinates {(\startx,\starty)};
                \addplot[mark=*, mark size=2pt, color=black, fill=green, forget plot] 
                    coordinates {(\endx,\endy)};
            }

        \addplot [thick, color=magenta, fill=magenta!60, circular legend] table {data/x0Data.dat};
        \addlegendentry{Obstacle};

        \plotTrajectory{data/xDataRCBFSim2.dat}{color=blue}{$x(t)$ (RLCBF)}
        \plotTrajectory{data/xHatDataRCBFSim2.dat}{dashed, color=cyan}{$\hat{x}(t)$ (RLCBF)}
        \plotTrajectory{data/xDataCBFSim2.dat}{color=purple!50!black}{$x(t)$ (LCBF)}
        \plotTrajectory{data/xHatDataCBFSim2.dat}{dashed, color=orange!85!black }{$\hat{x}(t)$ (LCBF)}
        \plotTrajectory{data/xDataNoCBFSim2.dat}{color=red}{$x(t)$ (No CBF)}
        \plotTrajectory{data/xHatDataNoCBFSim2.dat}{dashed, color=magenta}{$\hat{x}(t)$ (No CBF)}
    
        \plotMarkersOnly{data/xDataNoCBFSim2.dat}{red}
        \plotMarkersOnly{data/xHatDataNoCBFSim2.dat}{magenta}
    \end{axis}

\end{tikzpicture}

%% file: figures/statePlotSim1.tex
\begin{tikzpicture}
    \begin{axis}[
        xlabel={$t [s]$ },
        ylabel={$x(t)$},
        legend pos = north east,
        legend style={nodes={scale=0.5, transform shape}},
        enlarge y limits=0.05,
        enlarge x limits=0,
        width=\linewidth,
        height=0.4\linewidth,
    ]
    \pgfplotsinvokeforeach{1}{
        \addplot+ [mark=none] table [x index=0, y index=#1] {data/x1DataSim1.dat};
    }
    \pgfplotsinvokeforeach{1}{
        \addplot+ [ dashed, mark=none] table [x index=0, y index=#1] {data/x1HatDataSim1.dat};
    }

    \pgfplotsinvokeforeach{1}{
        \addplot+ [mark=none] table [x index=0, y index=#1] {data/x2DataSim1.dat};
    }
    \pgfplotsinvokeforeach{1}{
        \addplot+ [color = green!50!black,  dashed, mark=none] table [x index=0, y index=#1] {data/x2HatDataSim1.dat};
    }
    \legend{$(x)_{1}$, $(\hat{x})_{1}$, $(x)_{2}$, $(\hat{x})_{2}$}
    \end{axis}
\end{tikzpicture}

%% file: figures/criticWeightsPlotSim1.tex
\begin{tikzpicture}
    \begin{axis}[
        xlabel={$t [s]$},
        ylabel={$\hat{W}_{c}(t)$},
        legend pos = north east,
        legend style={nodes={scale=0.5, transform shape}},
        enlarge y limits=0.05,
        enlarge x limits=0,
        width=\linewidth,
        height=0.4\linewidth,
    ]
    \pgfplotsinvokeforeach{1,...,6}{
        \addplot+ [solid, color=color#1, mark=none] table [x index=0, y index=#1] {data/WcHatDataSim1.dat};
    }
    \legend{$(\hat{W}_{c})_{1}$, $(\hat{W}_{c})_{2}$, $(\hat{W}_{c})_{3}$,
    $(\hat{W}_{c})_{4}$, $(\hat{W}_{c})_{5}$, $(\hat{W}_{c})_{6}$}
    \end{axis}
\end{tikzpicture}

%% file: figures/controlPlotSim1.tex
\begin{tikzpicture}
    \begin{axis}[
        xlabel={$t [s]$},
        ylabel={$\hat{u}(t)$},
        legend pos = north east,
        legend style={nodes={scale=0.5, transform shape}},
        enlarge y limits=0.05,
        enlarge x limits=0,
        width=\linewidth,
        height=0.4\linewidth,
    ]
    \pgfplotsinvokeforeach{1}{
        \addplot+ [mark=none] table [x index=0, y index=#1] {data/uHatDataSim1.dat};
    }
    \legend{$\hat{u}$}
    \end{axis}
\end{tikzpicture}

%% file: figures/statePlotSim2.tex
\begin{tikzpicture}
    \begin{axis}[
        xlabel={$t [s]$},
        ylabel={$x(t)$},
        legend pos = north east,
        legend style={nodes={scale=0.5, transform shape}},
        enlarge y limits=0.05,
        enlarge x limits=0,
        width=\linewidth,
        height=0.4\linewidth,
    ]
    \pgfplotsinvokeforeach{1}{
        \addplot+ [mark=none] table [x index=0, y index=#1] {data/x1DataSim2.dat};
    }
    \pgfplotsinvokeforeach{1}{
        \addplot+ [ dashed, mark=none] table [x index=0, y index=#1] {data/x1HatDataSim2.dat};
    }

    \pgfplotsinvokeforeach{1}{
        \addplot+ [mark=none] table [x index=0, y index=#1] {data/x2DataSim2.dat};
    }
    \pgfplotsinvokeforeach{1}{
        \addplot+ [color=green!50!black,  dashed, mark=none] table [x index=0, y index=#1] {data/x2HatDataSim2.dat};
    }
    \legend{$(x)_{1}$, $(\hat{x})_{1}$, $(x)_{2}$, $(\hat{x})_{2}$}
    \end{axis}
\end{tikzpicture}

%% file: figures/criticweightsPlotSim2.tex
\begin{tikzpicture}
    \begin{axis}[
        xlabel={$t [s]$},
        ylabel={$\hat{W}_{c}(t)$},
        legend pos = north east,
        legend style={nodes={scale=0.5, transform shape}},
        enlarge y limits=0.05,
        enlarge x limits=0,
        width=\linewidth,
        height=0.4\linewidth,
    ]
    \pgfplotsinvokeforeach{1,...,6}{
        \addplot+ [solid, color=color#1, mark=none] table [x index=0, y index=#1] {data/WcHatDataSim2.dat};
    }
    \legend{$(\hat{W}_{c})_{1}$, $(\hat{W}_{c})_{2}$, $(\hat{W}_{c})_{3}$,
    $(\hat{W}_{c})_{4}$, $(\hat{W}_{c})_{5}$, $(\hat{W}_{c})_{6}$}
    \end{axis}
\end{tikzpicture}

%% file: figures/controlPlotSim2.tex
\begin{tikzpicture}
    \begin{axis}[
        xlabel={$t [s]$},
        ylabel={$\hat{u}(t)$},
        legend pos = north east,
        legend style={nodes={scale=0.5, transform shape}},
        enlarge y limits=0.05,
        enlarge x limits=0,
        width=\linewidth,
        height=0.4\linewidth,
    ]
    \pgfplotsinvokeforeach{1}{
        \addplot+ [ mark=none] table [x index=0, y index=#1] {data/uHatDataSim2.dat};
    }
    \legend{$\hat{u}$}
    \end{axis}
\end{tikzpicture}